\documentclass[aps,preprint,superscriptaddress,12pt]{revtex4-2}
\usepackage{graphicx}
\usepackage{amsmath}
\usepackage{amssymb}
\usepackage{float}
\usepackage{hyperref}% add hypertext capabilities
\usepackage{multirow}
\usepackage{media9}
\usepackage{animate}

\begin{document}

%\title{The response of graphene nanomechanical resonators to a digitally-modulated video signal}
%\title{Graphene nanomechanical vibrations driven by a digitally-modulated video signal}
%\title{Graphene nanomechanical vibrations driven by a video signal}
%\title{A graphene nanomechanical video receiver}
\title{A few-layer graphene nanomechanical resonator driven by digitally modulated video signals}
\author{Ce Zhang}
\author{Heng Lu}
\author{Chen Yang}
\author{YuBin Zhang}
\author{FengNan Chen}
\author{Ying Yan}
\author{Joel Moser}
\email{j.moser@suda.edu.cn}
\affiliation{School of Optoelectronic Science and Engineering \& Collaborative Innovation Center of Suzhou Nano Science and Technology, Soochow University, Suzhou, China}
\affiliation{Key Lab of Advanced Optical Manufacturing Technologies of Jiangsu Province \& Key Lab of Modern Optical Technologies of Education Ministry of China, Soochow University, Suzhou, China}
\date{}

\begin{abstract}
Nanomechanical resonators driven by multifrequency signals combine the physics of mesoscopic vibrations and the technologies of radio communication. Their simplest property stems from their resonant response: they behave as filters, responding only to driving signals whose frequency range is contained within that of the mechanical response. While the response is routinely probed with a single tone drive, a multifrequency drive offers the possibility of inducing richer vibrational dynamics. In this case, all the frequency components of the drive are simultaneously transduced into vibrations with different amplitudes and phases that superimpose and interfere. Here, we employ a few-layer graphene nanomechanical resonator as a filter for broadband, digitally modulated video signals. We transduce the modulated drive into modulated vibrations, which we demodulate into a nanomechanical video. Our work reveals distinctive features in vibrations driven by a coherent, multifrequency drive unseen in vibrations actuated by single tone drives or by a force noise.
\end{abstract}

\maketitle
\newpage

\section{Introduction}

Nanomechanical resonators enable the study of mesoscopic vibrations within individual systems that can be tuned and controlled \cite{Lemme2020,Xu2022,Bachtold2022,Schmid2023}. Owing to their small masses, they transduce weak driving forces near resonance into vibrations of sizeable amplitude. Their frequency response resembles that of RLC tuning circuits in analogue radio receivers, with the force playing the role of the driving voltage and the mechanical displacement playing the role of the circuit charge. Where the resonant frequency of the electrical response can be tuned with a variable capacitor or a variable inductor, the resonant frequencies of the mechanical response, which are the frequencies of the vibrational modes, can be adjusted in-situ by harnessing capacitive \cite{Nathanson1967,Sazonova2004} and dielectric forces \cite{Unterreithmeier2009}, and by straining the resonator either mechanically \cite{Goldsche2018} or through its thermal expansion \cite{Singh2010,Chiout2023}. As with the electrical response, the frequency bandwidth of the mechanical response is determined by dissipation and by resonant frequency fluctuations \cite{Bachtold2022}. A narrower electrical response enhances frequency selectivity, while a narrower mechanical response facilitates the sensing of smaller driving forces. A broader response is also advantageous. For example, a large-bandwidth tuning circuit is necessary to isolate a modulated radio signal and capture the entirety of its spectrum, while filtering away signals in adjacent bands. Similarly, nanomechanical resonators with a large bandwidth $\Delta f$ can respond to nearly resonant forces whose spectra are as broad as the mechanical response \cite{Stambaugh2006,Yaxing2015,Maillet2017,Huang2019}. They can also be used to study resonant frequency fluctuations induced much below resonance by fluctuators with correlation times as short as $1/\Delta f$ \cite{Dykman2010,Siria2012,Fong2012,Yaxing2014,Sun2015,Kalaee2019}. In essence, nanomechanical resonators behave as tunable bandpass filters for nearly resonant forces and as detectors for nonresonant fluctuators.

While the nanomechanical response to broadband force noise and low frequency fluctuators has been thoroughly studied, the response to multifrequency coherent drives has thus far received less attention. Where multifrequency signals are applied, oftentimes only a single frequency component is used to measure the response while the other components are used for different tasks. For example, amplitude modulated \cite{Sazonova2004,Chen2009,Nichol2012} and frequency modulated signals \cite{Gouttenoire2010} have been employed to drive the resonator at the single frequency of the carrier, while the modulation sidebands have been used to detect the vibrations. Two-tone signals, with one tone near resonance and the second tone outside of the response, have been used to study a variety of vibrational phenomena. Examples abound and include pump-probe spectroscopy, where intermode coupling is measured \cite{Westra2010}; parametrically-induced mode coupling, used to heat or cool vibrations \cite{Dykman1978,Mahboob2012,DeAlba2016,Mathew2016} and to engineer nonlinear friction \cite{Dong2018}; and symmetry-breaking parametric resonators used for binary information encoding \cite{Ryvkine2006,Mahboob2010,Leuch2016}. By contrast, the scope of applications for multifrequency signals whose entire spectra drive vibrations has been more limited. Two-tone drives have been used to demonstrate signal mixing with nonlinear nanomechanical resonators \cite{Erbe2000,Jaber2016}. The interplay of a two-tone drive near resonance and parametric pumps has enabled the realization of mechanical logic gates \cite{Mahboob2011}. The combination of a strong and a weak driving tone near resonance has been shown to squeeze thermomechanical noise \cite{Ochs2021}. Resonators driven nonlinearly by two large tones near resonance have displayed unconventional dynamics, with the beating between the tones acting as a slow modulation of the bifurcation points \cite{Houri2019,Catalini2024}. Overall, studying nanomechanical resonators using signals composed of multifrequency components sheds light on the physics of mesoscopic vibrations and enables useful applications.

Appealing demonstrations of the physics of multifrequency vibrations are provided by nanomechanical radio receivers and transmitters for frequency modulated (FM) signals \cite{Nguyen2009,Nguyen2013}. Nanomechanical FM receivers have been realized with suspended nanotubes \cite{Gouttenoire2010,Jensen2007,Rutherglen2007,Zhou2015} and with nanomachined silicon tuning forks \cite{Bartsch2012}. The FM voltage waveform is capacitively coupled to the receiving resonator, creating a modulated driving force that the resonator passively responds to. By contrast, nanomechanical FM transmitters ouput a modulated radio frequency signal. They have been demonstrated with suspended graphene \cite{Chen2013} and with nanomachined polysilicon disks \cite{Rocheleau2014,Naing2015}. In these devices, the vibrating element is part of an electrical positive feedback loop that drives it to a regime of self-sustaining oscillations. The frequency of the oscillations is modulated by a stream of low frequency voltage pulses or by an audio waveform, which creates a slowly varying capacitive force between the oscillator and a nearby electrode that slowly modulates the spring constant. Such a technique of `keying' the slowly modulated components, or baseband signal, of the vibrations is reminiscent of the use of a Morse key in telegraphy. In brief, nanomechanical radio receivers and transmitters embody the richness of mesoscopic vibrations driven by multifrequency drives. Importantly, they also offer the additional reward of an audible copy of the vibrations.

Here, we demonstrate that nanomechanical resonators based on few-layer graphene (FLG) can transduce a multifrequency, digital video signal into modulated vibrations that host an accurate copy of the video. We quantify the transduction using signal quality metrics for digital communications. We have chosen FLG resonators for their large resonant frequency tunabilities within the high frequency (HF) and the very high frequency (VHF) bands of the radio spectrum, and for the compromise they offer between broad bandwidths and moderate force sensitivities. Of interest in our study is the fact that the signal bandwidth is larger than or comparable to that of the mechanical response. This is in contrast to nanomechanical radio receivers and transmitters mentioned above, which process audio baseband signals whose bandwidth is necessarily much smaller than the mechanical bandwidth. A wide-bandwidth coherent drive acts as a useful probe to study the physics of the mechanical response. Because the resonator behaves as a filter, it shapes the multifrequency driving signal and transduces it into multifrequency vibrations with different phases and amplitudes that superimpose and interfere. We find that these interferences have a profound impact on the signal-to-noise ratio and on the bit error ratio (BER) of the transduction. We model the response to our multifrequency drive theoretically and are able to reproduce our data. We establish a relation between the mechanical bandwidth and the largest signal bandwidth that can be transduced without loss of information in the baseband components of the vibrations. Our work contributes to the recent research effort on the physics of resonators driven nonlinearly by multiple tones. Yet, where only two tones were used earlier \cite{Houri2019,Catalini2024}, here we study an unexplored regime where the response is coherently driven by a near continuum of frequency components within the whole mechanical bandwidth. We also envision experimental applications at a time of renewed interest in communications within the HF and the VHF bands \cite{Wang2018}. Even though their data rates are low, HF and VHF communications still provide the most robust and reliable means of long-distance communications in situations where access to other channels is compromised. Future HF communications will require the ability to occasionally transmit and receive weak signals encoded with compressed images and videos \cite{Wang2018}, while narrowband digital television has already been experimented with in the VHF band \cite{RSGB2024}. Our results show that nanomechanical resonators may act as passive receivers for such signals, as they do for radiowaves modulated at audio frequencies. 

\begin{figure}[b]
\centering
\includegraphics{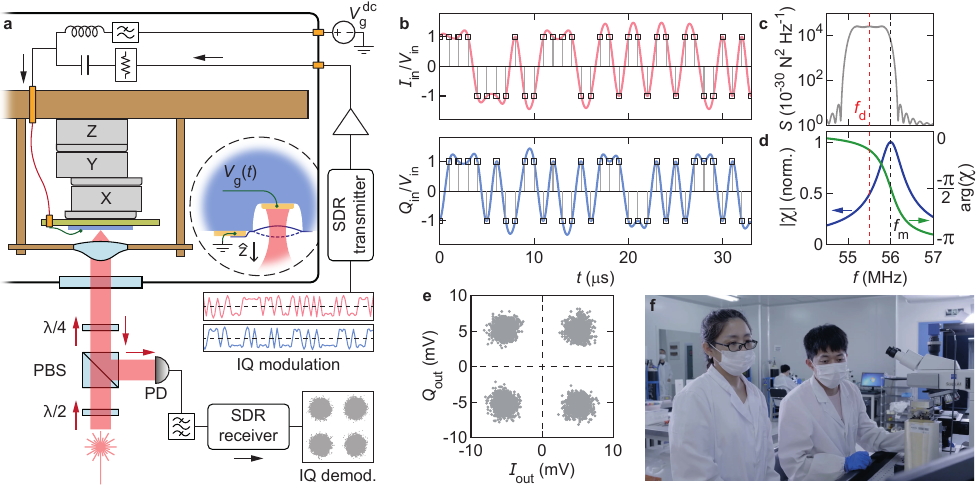}
\caption{\textbf{Principle of the experiment.} (a) Experimental setup. XYZ: piezo positioner. Red wavy line: flexible transmission line. $V_\mathrm{g}(t)=V_\mathrm{g}^\mathrm{dc}+s(t)$. PBS: polarizing beam splitter. $\lambda/2$, $\lambda/4$: half-, quarter-waveplate. An aspheric lens is used to focus incident light (wavelength $\simeq633$~nm) on the resonator area and to collect light reflected by the gate \cite{Haupt2014}. SDR: software-defined radio. `IQ modulation' represents two experimental streams of filtered voltage pulses forming one symbol stream. Black (red) arrows show the direction of propagation of radio frequency (optical) waves. (b) Computed streams of pulses before (vertical bars terminated by squares) and after (red and blue wavy traces) being convolved with a raised cosine (RC) filter. Symbol rate $R_\mathrm{sym}=10^6$~Hz. (c) Averaged power spectral density of driving force $S(f)$ as a function of spectral frequency calculated with a RC filter, for a drive power at the gate $P_\mathrm{d}=-31$~dBm, and for $R_\mathrm{sym}=10^6$~Hz. $S(f)$ is superimposed on the background of thermal forces. (d) Magnitude and argument of the linear mechanical response $\chi$ calculated for $Q_\mathrm{m}=100$. (e) Constellation diagram for $I_\mathrm{out}$ and $Q_\mathrm{out}$, the amplitudes of the in-phase and quadrature components of the decimated baseband signal at the output of the photodetector. (f) Frame of a nanomechanically processed video obtained from $I_\mathrm{out}$ and $Q_\mathrm{out}$.}\label{fig1}
\end{figure}

\section{Results}

We start by briefly describing the principle of our measurements. Our experimental setup is shown in Fig.~\ref{fig1}a. The FLG resonator, estimated to be composed of $\sim$~10 layers based on its optical contrast, is shaped as a 3~$\mu$m diameter drum suspended over a local gate electrode. It is kept at room temperature in a vacuum enclosure and is held by a piezo positioner. Two similar devices have been measured, yielding comparable results. Homemade flexible transmission lines fabricated from copper-clad ribbons of Kapton are used to send radio frequency signals to the gate without impeding the motion of the positioner \cite{Haupt2014,Lu2021}. With the resonator grounded, the voltage waveform incident on the gate, $s(t)$, creates a capacitive force $F_\mathrm{d}(t)\simeq (1+\Gamma)C^\prime V_\mathrm{g}^\mathrm{dc}s(t)$, where $\Gamma\simeq1$ is the reflection coefficient at the gate, $C^\prime$ is the derivative of the capacitance between the resonator and the gate with respect to displacement along the out-of-plane $\hat{z}$-direction (Fig.~\ref{fig1}a), $V_\mathrm{g}^\mathrm{dc}$ is a dc bias applied to the gate, and $t$ is time. Near resonance, $F_\mathrm{d}(t)$ drives flexural vibrations along $\hat{z}$. To measure them, we place the resonator in an optical standing wave formed between the gate and a quarter-wave plate \cite{Ce2024}, a standard configuration for studying two-dimensional resonators \cite{Lemme2020,Bunch2007,Barton2012,Davidovikj2016,Chen2023}. The amount of optical energy absorbed by the resonator from the standing wave is modulated at the frequency of the flexural vibrations. This results in oscillations in the output voltage of a photodetector placed in the reflected light path \cite{Barton2012}.

The originality of our setup lies in our technique to create a modulated drive and to demodulate the nanomechanical response. The baseband signal of the drive is built from a video in MP4 format that we convert into a video Transport Stream (TS). TS data are divided into two bit streams, which are used to map every symbol in the baseband signal with 2 bits. We implement a 4-quadrature amplitude modulation scheme (4-QAM, equivalent to quadrature phase shift keying). In this scheme, each symbol is represented by two quasi-simultaneous voltage pulses $I_\mathrm{in}=V_\mathrm{in}p$ and $Q_\mathrm{in}=V_\mathrm{in}q$, with $V_\mathrm{in}$ a voltage and $p=\pm1$, $q=\pm1$ (vertical bars in Fig.~\ref{fig1}b). We filter the pulses with a root raised cosine (RRC) filter, thereby creating two smoothly varying signals $\tilde{I}_\mathrm{in}(t)$ and $\tilde{Q}_\mathrm{in}(t)$. This filter serves the double purpose of minimizing intersymbol interference (in combination with a second RRC filter at the receiver) and presenting the resonator with a drive that is slowly modulated. Without it, the resonator would filter away much of the information encoded in the pulses. $\tilde{I}_\mathrm{in}(t)$ and $\tilde{Q}_\mathrm{in}(t)$ are used to modulate the amplitude of the in-phase and quadrature components of the driving waveform, $s(t)=\tilde{I}_\mathrm{in}(t)\cos(2\pi f_\mathrm{d}t)+\tilde{Q}_\mathrm{in}(t)\sin(2\pi f_\mathrm{d}t)$, where $f_\mathrm{d}$ is the drive frequency (Supplementary Note~B). Both baseband modulation (bit-to-symbol mapping) and carrier modulation of $s(t)$ are performed with GNU Radio, an open-source software development toolkit for signal processing \cite{GNU}. $s(t)$ is converted into an analog signal using an inexpensive software-defined radio transmitter (HackRF One by Great Scott Gadgets \cite{SDR}). The calculated power spectral density $S$ of the driving force $F_\mathrm{d}(t)$, averaged over many realizations of $F_\mathrm{d}(t)$, is shown in Fig.~\ref{fig1}c as a function of spectral frequency $f$. We consider a drive power at the gate $P_\mathrm{d}=10\,\mathrm{log}_{10}\left[(1+\Gamma)^2\langle s^2(t)\rangle/(50\times10^{-3})\right]\simeq-31$~dBm, where $\langle\cdot\rangle$ averages over time. We use $V_\mathrm{g}^\mathrm{dc}=13$~V and $C^\prime\simeq9\times10^{-10}$~F~m$^{-1}$ estimated with COMSOL. $S(f)$ is centered at $f_\mathrm{d}$ and its full spectral bandwidth reads $W\simeq1.3R_\mathrm{sym}$, where $R_\mathrm{sym}=10^6$~Hz is the symbol rate and 1.3 accounts for the roll-off factor of RRC. The area under $S(f)$ is the variance of $F_\mathrm{d}(t)$; it does not depend on $R_\mathrm{sym}$, and only depends on the amplitude of the drive. We also emphasize that $F_\mathrm{d}(t)$ is not a delta-correlated noise; it is instead a coherent waveform (Supplementary Note~B). Figure~\ref{fig1}d shows the normalized magnitude and the argument of the linear mechanical response. $W$ exceeds the bandwidth of the response $\Delta f=f_\mathrm{m}/Q_\mathrm{m}\simeq0.56\times10^6$~Hz, where $f_\mathrm{m}\simeq56\times10^6$~Hz is the resonant frequency of the fundamental mode and $Q_\mathrm{m}\simeq100$ is its quality factor. To measure the voltage $v(t)$ at the output of the photodetector, we employ an inexpensive software-defined radio receiver (NESDR Smart by Nooelec \cite{SDR}) that downconverts it to baseband. Using GNU Radio, we correct the received signal for the delay in transmission using a polyphase filter bank \cite{Franks1980,Harris2001} and we process it with another RRC filter (an operation known as matched filtering). We then obtain the in-phase and quadrature components of $v(t)$, which are defined as $v(t)=\tilde{I}_\mathrm{out}(t)\cos(2\pi f_\mathrm{d}t)+\tilde{Q}_\mathrm{out}(t)\sin(2\pi f_\mathrm{d}t)$. We decimate $\tilde{I}_\mathrm{out}(t)$ and $\tilde{Q}_\mathrm{out}(t)$ at the rate $R_\mathrm{sym}$ to obtain the symbol components $I_\mathrm{out}$ and $Q_\mathrm{out}$. These components, acquired for 2~s, are plotted on a constellation diagram in Fig.~\ref{fig1}e. This diagram helps visualize the process of converting symbols back into bits, the decision to ascribe a received symbol to a pair of bits depending on which quadrant the symbol appears in \cite{Crockett2023}. The GNU Radio flowcharts used for these operations are shown in Supplementary Note~C. Figure~\ref{fig1}f displays a frame from the video obtained by converting the streams of $I_\mathrm{out}$ and $Q_\mathrm{out}$ into a video in MP4 format using FFmpeg, an open-source software for processing multimedia files \cite{FFmpeg}. It portrays two of the authors (Y. Y. and Y. B. Z.) in our lab. The nanomechanically processed video is available as Supplementary Video~1.

\begin{figure}[t]
\centering
\includegraphics{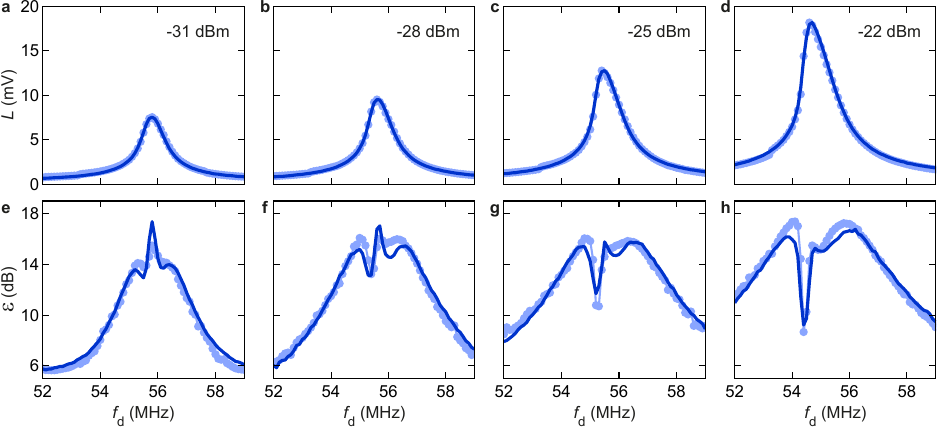}
\caption{\textbf{Nanomechanical response to a drive modulated with a video signal at $R_\mathrm{sym}=0.5\times10^6$~Hz.} (a-d) Vector length $L$ of $I_\mathrm{out}$ and $Q_\mathrm{out}$ as a function of drive frequency $f_\mathrm{d}$. (e-h) Reciprocal of the error vector magnitude $\varepsilon(f_\mathrm{d})$. Drive power at the gate $P_\mathrm{d}=-31$~dBm (a, e), $-28$~dBm (b, f), $-25$~dBm (c, g), and $-22$~dBm (d, h). Dots are measured data. Solid lines are calculated traces. $V_\mathrm{g}^\mathrm{dc}=13$~V in all panels.}\label{fig2}
\end{figure}

Obtaining a nanomechanical video requires optimizing $P_\mathrm{d}$, $f_\mathrm{d}$ and $R_\mathrm{sym}$. We do so by changing these parameters one by one and monitoring the effect on the constellation diagram for $I_\mathrm{out}$ and $Q_\mathrm{out}$. As $f_\mathrm{d}$ is stepped through resonance, this diagram experiences significant changes that are best visualized as animations (Supplementary Video~2 shows constellation diagrams measured upon increasing $f_\mathrm{d}$, at different values of $P_\mathrm{d}$, and at $R_\mathrm{sym}=0.5\times10^6$~Hz). The simplest quantity we extract from the constellation diagram is the time-averaged length of the $I-Q$ vector in one quadrant, $L=\langle\vert I_\mathrm{out} + \mathrm{i}Q_\mathrm{out}\vert\rangle$. The spread of the symbol clouds (see Fig.~\ref{fig1}e) is informative as well. We define it within one quadrant as $d=\langle(I_\mathrm{out}-\langle I_\mathrm{out}\rangle)^2+(Q_\mathrm{out}-\langle Q_\mathrm{out}\rangle)^2\rangle^{1/2}$. One may interpret the logarithm of the ratio, $\varepsilon=20\,\mathrm{log}_{10}(L/d)$, as a signal-to-noise ratio, because noise at the output of our receiver and the dephasing of vibrations by the mechanical response both contribute to the spread of the clouds (see below). In digital communications, $\varepsilon$ is known as the reciprocal of the error vector magnitude of the symbols. The larger $\varepsilon$ is, the more separated the symbol clouds are and the smoother the video playback will be. Measured $L(f_\mathrm{d})$ and $\varepsilon(f_\mathrm{d})$ data are shown as dots in Fig.~\ref{fig2}a-d and Fig.~\ref{fig2}e-h, respectively, for several values of $P_\mathrm{d}$ and at $R_\mathrm{sym}=0.5\times10^6$~Hz (the smallest $L$ and $\varepsilon$ for which a video can be decoded are shown in Supplementary Note~D). They are similarly shown in Figs.~\ref{fig3}a-d and Figs.~\ref{fig3}e-h for $R_\mathrm{sym}=10^6$~Hz. We have set $V_\mathrm{g}^\mathrm{dc}=13$~V, corresponding to $f_\mathrm{m}\simeq56\times10^6$~Hz. For both values of $R_\mathrm{sym}$, $L(f_\mathrm{d})$ broadens and becomes more asymmetric as $P_\mathrm{d}$ increases, a behavior indicative of nonlinearities in the response. While $L(f_\mathrm{d})$ traces simply resemble broad resonances, $\varepsilon(f_\mathrm{d})$ traces contain more features. At low $P_\mathrm{d}$, $\varepsilon(f_\mathrm{d})$ peaks near $f_\mathrm{m}$ and develops two local minima off resonance (Fig.~\ref{fig2}e), the latter being more pronounced at $R_\mathrm{p}=10^6$~Hz (Fig.~\ref{fig3}e). At high $P_\mathrm{d}$, the local minimum in $\varepsilon(f_\mathrm{d})$ at $f_\mathrm{d}$ where $\vert\partial L/\partial f_\mathrm{d}\vert$ is largest turns into a deep minimum, and the maxima in $\varepsilon(f_\mathrm{d})$ shift away from resonance (Figs.~\ref{fig2}h, \ref{fig3}h). These maxima and minima in $\varepsilon(f_\mathrm{d})$ are important features because they impact the quality of the demodulated video.

\begin{figure}[t]
\centering
\includegraphics{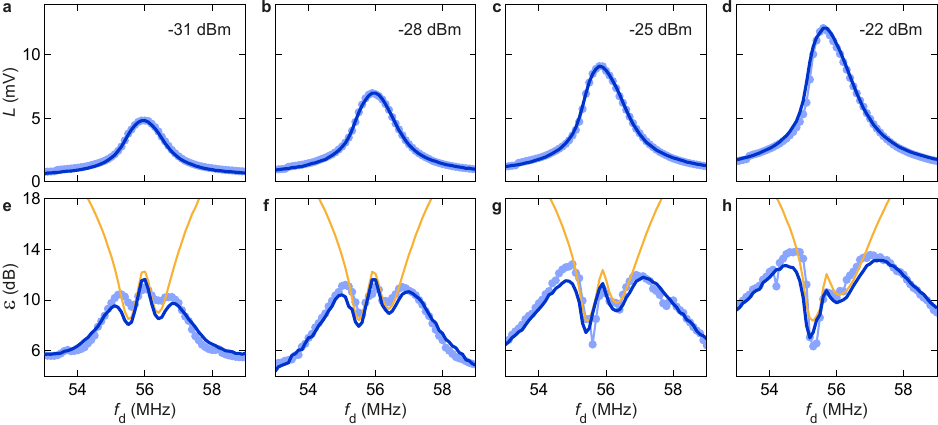}
\caption{\textbf{Nanomechanical response to a drive modulated with a video signal at $R_\mathrm{sym}=10^6$~Hz.} (a-d) Vector length $L$ of $I_\mathrm{out}$ and $Q_\mathrm{out}$ as a function of drive frequency $f_\mathrm{d}$. (e-h) Reciprocal of the error vector magnitude $\varepsilon(f_\mathrm{d})$. Drive power at the gate $P_\mathrm{d}=-31$~dBm (a, e), $-28$~dBm (b, f), $-25$~dBm (c, g), and $-22$~dBm (d, h). Dots are measured data [measurement noise in (f) is a little higher than in (e, g, h)]. Blue (orange) solid lines are calculated traces with noise on (off). $V_\mathrm{g}^\mathrm{dc}=13$~V in all panels.}\label{fig3}
\end{figure}

\section{Discussion}

To understand these observations, we numerically solve the equation of motion for the resonator in response to the multifrequency driving force $F_\mathrm{d}(t)$. The equation reads \cite{Dykman1975,Zaitsev2012}:
\begin{equation}
\ddot{z} + \left[\frac{\omega_\mathrm{m}}{Q_\mathrm{m}}+\frac{\eta}{m_\mathrm{eff}}z^2\right]\dot{z} + \left[\omega_\mathrm{m}^2+\frac{\alpha}{m_\mathrm{eff}}z^2\right]z=\frac{F_\mathrm{d}(t)}{m_\mathrm{eff}}\,,\label{nld}
\end{equation}
with $z$ the resonator displacement, $\omega_\mathrm{m}=2\pi f_\mathrm{m}$ the angular resonant frequency, $\eta$ the nonlinear damping coefficient, and $\alpha$ the Duffing elastic constant. $Q_\mathrm{m}\simeq100$, $\eta\simeq1.2\times10^5$~kg~m$^{-2}$~s$^{-1}$ and $\alpha\simeq-10^{14}$~kg~m$^{-2}$~s$^{-2}$ are estimated from the response to a single tone drive (Supplementary Note~E). $m_\mathrm{eff}\simeq2\times10^{-17}$~kg is estimated from the shape of the resonator and the number of graphene layers. As in the experiment, $\tilde{I}_\mathrm{in}(t)$ and $\tilde{Q}_\mathrm{in}(t)$ that define the baseband signal of $F_\mathrm{d}(t)$ are computed from two streams of pulses. The height of the pulses takes on the values $\pm V_\mathrm{in}$ randomly at a rate $R_\mathrm{sym}$. Pulse streams are processed with a raised cosine filter (Fig.~\ref{fig1}b). The use of this filter is justified in the case where most of the experimental noise appears at the output of the receiver, as in our measurements. For completeness, we have also considered the case where, in addition to noise at the receiver, a white Gaussian force noise is added to the drive. This case requires a cascade of two RRC filters, one at the transmitter and one at the receiver. Yet, we have found that the simpler scenario based on receiver noise only is sufficient to explain our measurements. Demodulating $z(t)$ yields $\tilde{I}_\mathrm{out}(t)=2h(t)\ast[\xi z(t)\cos(2\pi f_\mathrm{d}t)]$ and $\tilde{Q}_\mathrm{out}(t)=2h(t)\ast[\xi z(t)\sin(2\pi f_\mathrm{d}t)]$, where $h$ is a lowpass filter, $\xi$ is a conversion factor in units of V~m$^{-1}$, and $\ast$ is the convolution operator. We obtain the symbol components $I_\mathrm{out}$ and $Q_\mathrm{out}$ by decimating $\tilde{I}_\mathrm{out}(t)$ and $\tilde{Q}_\mathrm{out}(t)$ at the rate $R_\mathrm{sym}$, taking into account the delay between input and output signals caused by the raised cosine filter. We find that this simple model reproduces our measured $L(f_\mathrm{d})$ and $\varepsilon(f_\mathrm{d})$ data (calculations are shown as solid blue traces in Figs.~\ref{fig2} and \ref{fig3}). The only fit parameter in our analysis of $L(f_\mathrm{d})$ is $\xi\simeq7\times10^5$~V~m$^{-1}$. The lineshape of $L(f_\mathrm{d})$ is broader than the response to a drive with unmodulated in-phase and quadrature components at the same $P_\mathrm{d}$ (Supplementary Note~B). To analyze $\varepsilon(f_\mathrm{d})$, we model fluctuations $\delta\tilde{I}_\mathrm{out}$ and $\delta\tilde{Q}_\mathrm{out}$ at the output of the receiver as additive white Gaussian noise with variance $\sigma^2$. This noise contributes to the spread of our calculated symbol clouds as $d^2\rightarrow d^2+\langle\delta\tilde{I}_\mathrm{out}^2\rangle+\langle\delta\tilde{Q}_\mathrm{out}^2\rangle=d^2+2\sigma^2$. We use $\sigma^2\simeq3.9\times10^{-19}$~m$^2$ as a single fit parameter in our analysis of $\varepsilon(f_\mathrm{d})$. Measurements of the noise at the output of the receiver are shown in Supplementary Note~F.

Interestingly, the lineshape of $\varepsilon(f_\mathrm{d})$ near $f_\mathrm{m}$ is not influenced by the noise at the receiver. $\sigma^2$ is only needed to capture the decay of $\varepsilon(f_\mathrm{d})$ away from $f_\mathrm{m}$, but it has only little incidence on the most interesting features that are the minima in $\varepsilon(f_\mathrm{d})$. To show this, in Figs.~\ref{fig3}e-h we display our calculations for $\varepsilon(f_\mathrm{d})$ with no noise added, that is, without any fit parameter (orange traces). In this case, $\varepsilon$ increases as $f_\mathrm{d}$ is stepped away from $f_\mathrm{m}$, while the features near $f_\mathrm{m}$ are similar to those with noise on. This observation indicates that the spread of the symbol clouds $d(f_\mathrm{d})$ near $f_\mathrm{m}$ is caused by a process other than noise. This process is the loss of coherence of the response caused by the superposition of multifrequency vibrations with different amplitudes and phases. It is a distinctive feature in vibrations driven by a coherent, multifrequency drive that is unseen in vibrations actuated by single tone drives or by a force noise. It is a result of the resonator acting as a filter in the case where the spectrum of the drive is broader than the response, $W\gtrsim\Delta f$. If the drive were narrowband, $W\ll\Delta f$, then the symbol components $(I_\mathrm{in},Q_\mathrm{in})$ would simply be transformed into $(I_\mathrm{out},Q_\mathrm{out})$ by a rotation through the phase angle of the response. By contrast, every frequency component in the broadband drive gets transduced into a vibrational frequency component with a different amplitude and a different phase. Where $f_\mathrm{d}$ is far away from $f_\mathrm{m}$, and provided that receiver noise is low, $(I_\mathrm{out},Q_\mathrm{out})$ still result from a simple rotation through a single angle ($\simeq0$ or $-\pi$), since the magnitude and the phase of the response depend on frequency only weakly. In this case, $I_\mathrm{out}\propto\pm I_\mathrm{in}$ and $Q_\mathrm{out}\propto\pm Q_\mathrm{in}$. Where $f_\mathrm{d}\simeq f_\mathrm{m}$, $W\simeq\Delta f$, and where the response is linear, we find $I_\mathrm{out}\propto Q_\mathrm{in}$ and $Q_\mathrm{out}\propto-I_\mathrm{in}$ (Supplementary Note~G), so the symbols in the output stream are simply exchanged (with no incidence on the video demodulation, as shown in Supplementary Note~H). However, the dephasing effect of the response on $I_\mathrm{out}$ and $Q_\mathrm{out}$ is strongest where $\vert f_\mathrm{d}-f_\mathrm{m}\vert\simeq W/2$ and $W\gtrsim\Delta f$. This case is depicted in Figs.~\ref{fig1}c-d, where the upper frequency edge of the drive spectrum is nearly resonant while the lower frequency edge is away from resonance. There, we find that $\tilde{I}_\mathrm{out}(t)$ contains dephased frequency components both from $\tilde{I}_\mathrm{in}(t)$ and from $\tilde{Q}_\mathrm{in}(t)$, and so does $\tilde{Q}_\mathrm{out}(t)$ [Supplementary Note~G]. As a result, the decimation of $\tilde{I}_\mathrm{out}(t)$ and $\tilde{Q}_\mathrm{out}(t)$ produces broad distributions of $I_\mathrm{out}$ and $Q_\mathrm{out}$ going through zero, which decreases $\varepsilon$. Experimentally, $\vert f_\mathrm{d}-f_\mathrm{m}\vert\simeq W/2$ is indeed where we find minima in $\varepsilon(f_\mathrm{d})$ for $R_\mathrm{sym}=0.5\times10^6$~Hz and $10^6$~Hz in the linear regime (low $P_\mathrm{d}$).

In the nonlinear regime (high $P_\mathrm{d}$), our model also reproduces the deepening of the local minimum (dip) in $\varepsilon(f_\mathrm{d})$ near the largest slope in $L(f_\mathrm{d})$ and the frequency shift of the dip $\Delta f_\mathrm{dip}$ as $P_\mathrm{d}$ is increased. We show in Supplementary Note~I that $\Delta f_\mathrm{dip}$ is nearly proportional to the square of the change in vector length $\Delta L_\mathrm{dip}$, $\Delta f_\mathrm{dip}\propto\Delta L_\mathrm{dip}^2$. This behavior is consistent with the quartic relation between the frequency of a vibrational mode and the vibrational amplitude caused by weak conservative nonlinearities in the case where the mode is driven by a single tone drive \cite{Carr1999,Husain2003,Kozinsky2006}. It is also consistent with the lineshape broadening of the mechanical response induced by the interplay between conservative nonlinearities and a force noise \cite{Dykman1971,Maillet2017,Huang2019}, as described in Supplementary Note~I.

\newpage

\begin{figure}[H]
\centering
\includegraphics{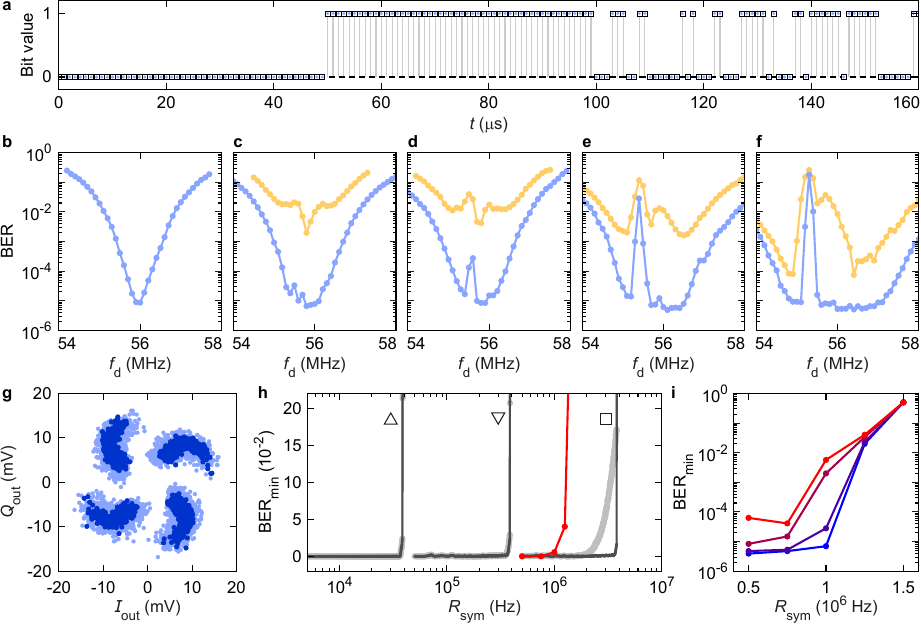}
\caption{\textbf{Bit error ratio (BER) of the nanomechanical transduction.} (a) Bit stream at the transmitter (hollow squares) and bit stream at the receiver (dots) measured at $R_\mathrm{sym}=0.5\times10^6$~Hz. Two bits make one symbol in 4-QAM, so the bit rate is twice the symbol rate. (b-f) BER as a function of drive frequency $f_\mathrm{d}$ at $R_\mathrm{sym}=0.5\times10^6$~Hz (blue dots) and at $R_\mathrm{sym}=10^6$~Hz (orange dots) for drive power $P_\mathrm{d}=-34$~dBm (b), $-31$~dBm (c), $-28$~dBm (d), $-25$~dBm (e), and $-22$~dBm (f). (g) Constellation diagram for $I_\mathrm{out}$ and $Q_\mathrm{out}$ at $P_\mathrm{d}=-25$~dBm with $f_\mathrm{d}$ set to the BER peak frequency in (e). Light (dark) blue dots are measurements (calculations). (h) Minimum of BER in the linear regime as a function of $R_\mathrm{sym}$. Grey traces are calculated without (dark grey) and with added noise at the receiver (light grey). Up triangle: $\Delta f=10^4$~Hz. Down triangle: $\Delta f=10^5$~Hz. Square: $\Delta f=10^6$~Hz. Red dots show measurements at $P_\mathrm{d}=-28$~dBm. (i) Measured $\mathrm{BER}_\mathrm{min}$ as a function of $R_\mathrm{sym}$. From top (red) to bottom (blue), $P_\mathrm{d}=-28$~dBm, $-25$~dBm, $-22$~dBm, $-19$~dBm. Measurements in (h, i) were performed in a noisier environment than measurements shown in (b-g). $V_\mathrm{g}^\mathrm{dc}=13$~V in all panels.}\label{fig4}
\end{figure}

To complement our measurements of $\varepsilon$ and directly quantify the effect of the mechanical response on the driving bit stream, we measure the bit error ratio ($\mathrm{BER}$) of the nanomechanical transduction. The latter is the ratio of the number of misread bits in the received stream (output bit stream) to the number of bits in the stream sent by the transmitter (input bit stream). Measuring $\mathrm{BER}$ is not straighforward because propagation and filtering delays make it difficult to compare the two bit streams. We solve this problem with the help of a marker placed at the beginning of the input bit stream. The marker consists of a train of bits `0' followed by a train of bits `1'. It can be easily identified on the receiver side, allowing us to compare input and output bits one by one. As an example, Fig.~\ref{fig4}a displays a short input bit stream (hollow squares) and the received output bit stream (dots) obtained with $R_\mathrm{sym}=0.5\times10^6$~Hz. It is an example of a (short) error-free transduction, $\mathrm{BER}=0$. Figures~\ref{fig4}b-f show $\mathrm{BER}$ as a function of $f_\mathrm{d}$ measured at $R_\mathrm{p}=0.5\times10^6$~Hz (blue dots) and at $R_\mathrm{p}=10^6$~Hz (orange dots) from the linear regime to the nonlinear one [$P_\mathrm{d}$ increases from (b) to (f)]. At each $f_\mathrm{d}$, $\mathrm{BER}$ is computed from a single measurement of an output bit stream acquired for 15~s. We find that $\mathrm{BER}$ at $R_\mathrm{sym}=0.5\times10^6$~Hz is always lower than at $R_\mathrm{sym}=10^6$~Hz, indicating that $\mathrm{BER}_\mathrm{min}$, the minimum of $\mathrm{BER}$, depends on $R_\mathrm{sym}$. At the largest $P_\mathrm{d}$ and at $R_\mathrm{sym}=0.5\times10^6$~Hz (Fig.~\ref{fig4}f), we measure $\mathrm{BER}_\mathrm{min}\simeq5\times10^{-6}$, corresponding to 5 misread bits for every million bits sent to the resonator. It is limited by noise in our setup (Supplementary Notes~D and F). Further, we find that $\mathrm{BER}(f_\mathrm{d})$ develops a sharp peak as $P_\mathrm{d}$ is increased that is concomitant with the appearance of the lower frequency dip in $\varepsilon(f_\mathrm{d})$ (Supplementary Note~I). At the peak frequency, symbol clouds on the constellation diagram are strongly distorted, as confirmed by our model (Fig.~\ref{fig4}g), and the decoding of the nanomechanical video fails. In Supplementary Note~J, we show our calculated $L(f_\mathrm{d})$, $\varepsilon(f_\mathrm{d})$, $\mathrm{BER}(f_\mathrm{d})$ and constellation diagram as an embedded animation to easily appreciate how these quantities relate to one another.

With regard to applications, we now consider the dependence of $\mathrm{BER}_\mathrm{min}$ on $R_\mathrm{sym}$. We wish to find out how $R_\mathrm{sym}^\ast$, the largest $R_\mathrm{sym}$ at which $\mathrm{BER}_\mathrm{min}$ is small, changes with the mechanical linewidth $\Delta f$. We identify $R_\mathrm{sym}^\ast$ as the filtering bandwidth of the resonator for the video signal. We focus on $\mathrm{BER}$ at resonance in the linear regime, where $\mathrm{BER}(f_\mathrm{d})=\mathrm{BER}_\mathrm{min}$ (Fig.~\ref{fig4}b). In Fig.~\ref{fig4}h, we plot $\mathrm{BER}_\mathrm{min}$ as a function of $R_\mathrm{sym}$ calculated using Eq.~(\ref{nld}) for $\Delta f=10^4$~Hz, $10^5$~Hz, and $10^6$~Hz (grey traces from left to right). Dark traces are calculated with no noise added, while light traces are calculated with additive Gaussian noise of variance $\sigma^2$ used in Figs.~\ref{fig2} and~\ref{fig3}. Every trace without noise shows a kink near $R_\mathrm{sym}\simeq3.5\Delta f$, which marks an upper bound for $R_\mathrm{sym}^\ast$. Additive noise restricts $R_\mathrm{sym}^\ast$ to lower values, especially for larger $\Delta f$ (lower $Q_\mathrm{m}$), as $\mathrm{BER}_\mathrm{min}$ starts to increase from a lower $R_\mathrm{sym}$. Red dots in Fig.~\ref{fig4}h represent $\mathrm{BER}_\mathrm{min}(R_\mathrm{sym})$ measured at $P_\mathrm{d}=-28$~dBm, at the onset of the nonlinear regime, and in the presence of noise. We extract the experimental filtering bandwidth $R_\mathrm{sym}^\ast\simeq10^6$~Hz. $\mathrm{BER}_\mathrm{min}$ data measured over a range of $P_\mathrm{d}$ are displayed in Fig.~\ref{fig4}i (see also Supplementary Note~K). They show, somewhat expectedly, that nanomechanical information encoding incurs less information loss at higher drive power.

\begin{figure}[b]
\centering
\includegraphics{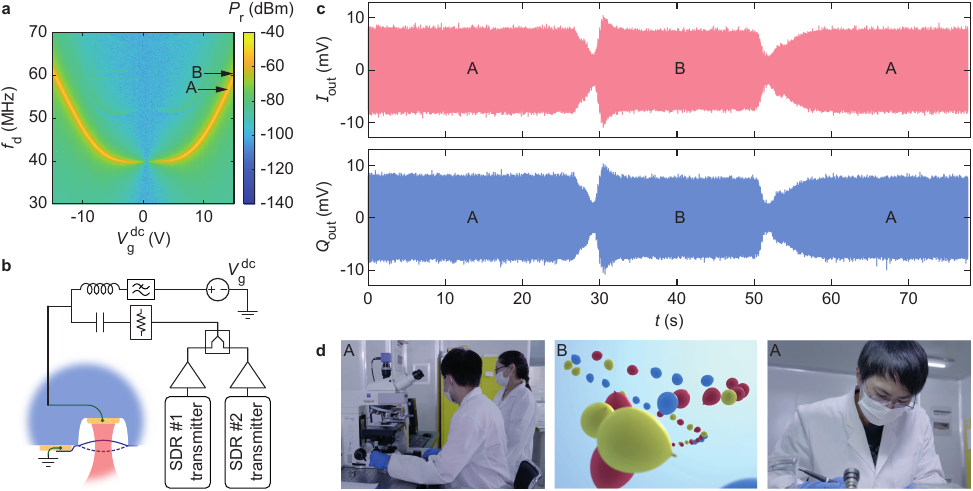}
\caption{\textbf{Multichannel nanomechanical television broadcast.} (a) Response of the resonator to a single tone drive as a function of gate voltage $V_\mathrm{g}^\mathrm{dc}$ and drive frequency $f_\mathrm{d}$ for a drive power of $-40$~dBm. (b) Experimental setup featuring two software-defined radios transmitting in parallel (a single receiver is used). (c) $\tilde{I}_\mathrm{out}(t)$ and $\tilde{Q}_\mathrm{out}(t)$ of the baseband component of the output signal. Segment A, $0\,\mathrm{s}<t<27\,\mathrm{s}$: $V_\mathrm{g}^\mathrm{dc}=14$~V and central frequency of the receiver $f_\mathrm{r}=57.8\times10^6$~Hz. Segment B, $29\,\mathrm{s}<t<50\,\mathrm{s}$: $V_\mathrm{g}^\mathrm{dc}=15$~V and $f_\mathrm{r}=60.8\times10^6$~Hz. Segment A, $t>52\,\mathrm{s}$: $V_\mathrm{g}^\mathrm{dc}=14$~V and $f_\mathrm{r}=57.8\times10^6$~Hz. (d) Frames of the videos being transmitted.}\label{fig5}
\end{figure}

We now demonstrate the filtering and the video encoding capabilities of our resonator. Figure~\ref{fig5}a shows that the nanomechanical response (measured here with a single tone drive) depends on $V_\mathrm{g}^\mathrm{dc}$. Given that $\varepsilon$ and $\mathrm{BER}$ strongly depend on $\vert f_\mathrm{m}-f_\mathrm{d}\vert$, we can send multiple videos with well separated $f_\mathrm{d}$ simultaneously and choose to demodulate one of them by properly setting $f_\mathrm{m}$ with $V_\mathrm{g}^\mathrm{dc}$. Our experimental setup for a two-video broadcast is depicted in Fig.~\ref{fig5}b. Two software-defined radio transmitters are used in parallel, one with $f_\mathrm{d}=57.8\times10^6$~Hz~$=f_\mathrm{A}$ and the other with $f_\mathrm{d}=60.8\times10^6$~Hz~$=f_\mathrm{B}$. Both transmitters are set with $R_\mathrm{p}=10^6$~Hz and $P_\mathrm{d}=-22$~dBm, at which power the resonator responds nonlinearly [both drive frequencies are set away from the dip in $\varepsilon(f_\mathrm{d})$]. Figure~\ref{fig5}c shows $\tilde{I}_\mathrm{out}(t)$ and $\tilde{Q}_\mathrm{out}(t)$, having set $V_\mathrm{g}^\mathrm{dc}=14$~V and the central frequency of the receiver $f_\mathrm{r}=f_\mathrm{A}$ at $t=0$. Between $t=27$~s and $t=29$~s, $V_\mathrm{g}^\mathrm{dc}$ is ramped up from 14~V to 15~V, and $f_\mathrm{r}$ is set to $f_\mathrm{B}$ in software. Between $t=50$~s and $t=52$~s, $V_\mathrm{g}^\mathrm{dc}$ is ramped back down from 15~V to 14~V, and $f_\mathrm{r}$ is set back to $f_\mathrm{A}$. Figure~\ref{fig5}d displays frames from the videos being transmitted, showing that we can conveniently tune in to the video broadcast of our choosing with $V_\mathrm{g}^\mathrm{dc}$. The full nanomechanical video is available as Supplementary Video~3.

In the spirit of previous nanomechanical devices employed as signal processors \cite{Nguyen2009}, we envision that our FLG resonators, and other resonators based on similar principles, may be employed as small footprint, passive radio receivers in the HF and the VHF bands of the radio spectrum. With a resonant frequency ranging from $\simeq40\times10^6$~Hz to $\simeq60\times10^6$~Hz (Fig.~\ref{fig5}a), our resonator operates in the 6-meter subband that is internationally allocated to the use of amateur radio. In this subband, digital amateur television (DATV) broadcasting has been conducted using signals with a 300~kHz bandwidth \cite{RSGB2024}. In the United Kingdom, DATV with a broader bandwidth of up to 500~kHz is allowed in the 2-meter amateur radio band (144--148~MHz) \cite{RSGB2024}. Smaller diameter FLG resonators, or higher order vibrational modes, can operate in this band. The filtering bandwidth of our resonator, and the maximum 4-QAM data rate of its transduction $2R_\mathrm{sym}^\ast\simeq2\times10^6$~bps, are then fully compatible with DATV. Arguably, our simple experimental setup may still not be practical enough for receiving amateur radio messages, but it can be simplified further \cite{Lu2021}. The resonator would serve the dual purpose of a front-end, tunable bandpass filter and of a passive optocoupler. It would be placed at the ouput of the preamplifier that is usually connected to the feedpoint of a receiving antenna \cite{ARRL2019}. As a filter, it would reject jamming and other interfering signals the way surface acoustic wave filters do, but with the additional benefit of a tunable center frequency, and without the fabrication complexity of more advanced tunable front-end filters \cite{Hashimoto2015}. As an optocoupler, it would passively convert the radio signal into modulations of reflected light intensity through the nanomechanical transduction, effectively eliminating the problem of radio frequency interferences caused by common-mode currents in the coaxial cable that connects the antenna to the demodulator \cite{ARRL2019}. Admittedly, the idea of employing nanomechanical resonators as devices for radio communcation mostly possesses an academic appeal, given the limitations posed by device reproducibility and thermal drifts \cite{Nguyen2013}. However, these resonators may ultimately enable alternate architectures for shortwave radio receivers, facilitating the decoding of images and videos carried by weak signals through a noisy environment \cite{Crockett2023}.

\section*{Data availability}
The data that support the findings of this study are available at \url{https://doi.org/10.6084/m9.figshare.28137728.v1}

\section*{Code availability}
The codes used in this study have been deposited in this repository: \url{https://github.com/Changchang2019/Nanomechanical_Videos}

\section*{Acknowledgement}

This work was supported by the National Natural Science Foundation of China (grant numbers 62150710547 and 62074107) and the project of the Priority Academic Program Development (PAPD) of Jiangsu Higher Education Institutions. The authors are grateful to Prof. Wang Chinhua for his strong support. J.M. is grateful to Dr. Martin Kroner for his advice on the optical measurement setup and on the flexible transmission lines.

\section*{Author contributions}
C.Z. and J.M. assembled the data acquisition system. H.L., C.Y. and Y.Z. fabricated the nanomechanical devices with the help of F.C. H.L. built the optical setup with the help of Y.Y. C.Z. performed the measurements. J.M. analyzed the data, wrote the manuscript and supervised the work.

\section*{Supplementary Material}

\subsection{Supplementary Videos}

Supplementary Videos accompany this work. They have been deposited in this repository: \url{https://doi.org/10.6084/m9.figshare.28494875}

\subsection{Driving voltage and driving force waveforms}\label{section_waveforms}

To build the driving voltage and the driving force waveforms, we start from two streams of pulses $p_k=\pm1$ and $q_k=\pm1$ with pulse (symbol) rate $R_\mathrm{sym}$. The streams are filtered either with a raised cosine (RC) filter or with a root raised cosine (RRC) filter, yielding waveforms $\tilde{p}(t)$ and $\tilde{q}(t)$ that vary continuously in time $t$. Where noise is added to the signal after the transmitter and before the receiver, a RRC filter is used to filter the pulses on the transmitter side and another is used to filter the baseband signal on the receiver side. A cascade of two RRC filters minimizes intersymbol interference, as does a single RC filter without noise. $\tilde{p}(t)$ and $\tilde{q}(t)$ can be expressed as:
\begin{equation}
\tilde{p}(t)=\sum_{k=0}^{\infty}p_k\,g(t-kT_\mathrm{sym})\,,\quad\tilde{q}(t)=\sum_{k=0}^{\infty}q_k\,g(t-kT_\mathrm{sym})\,,
\end{equation}
where $g$ is the impulse response of the filter and $T_\mathrm{sym}$ is the reciprocal of the symbol rate, $T_\mathrm{sym}=1/R_\mathrm{sym}$. For a RC filter, $g$ reads:
\begin{eqnarray}
g(t)&=&\frac{\pi}{4T_\mathrm{sym}}\mathrm{sinc}\left(\frac{1}{2\beta}\right)\quad\mathrm{if}\quad t=\pm\frac{T_\mathrm{sym}}{2\beta}\\
g(t)&=&\frac{1}{T_\mathrm{sym}}\mathrm{sinc}\left(\frac{t}{T_\mathrm{sym}}\right)\frac{\cos\left(\frac{\pi\beta t}{T_\mathrm{sym}}\right)}{1-\left(\frac{2\beta t}{T_\mathrm{sym}}\right)^2}\quad\mathrm{otherwise,}
\end{eqnarray}
where $\beta$ is the roll-off factor of the filter. The filtered pulses are used to modulate the amplitude of the in-phase component and that of the quadrature component of the waveform $x(t)$ defined as:
\begin{equation}
x(t)=\tilde{p}(t)\cos(\omega_\mathrm{d}t)+\tilde{q}(t)\sin(\omega_\mathrm{d}t)\,,\label{x}
\end{equation}
where $\omega_\mathrm{d}$ is the angular frequency of the drive. Given a peak voltage amplitude $V_\mathrm{in}$, the driving voltage waveform $s(t)$ is defined as:
\begin{equation}
s(t)=V_\mathrm{in}x(t)=\tilde{I}_\mathrm{in}(t)\cos(\omega_\mathrm{d}t)+\tilde{Q}_\mathrm{in}(t)\sin(\omega_\mathrm{d}t)\,,\label{s}
\end{equation}
where $\tilde{I}_\mathrm{in}(t)$ and $\tilde{Q}_\mathrm{in}(t)$ are the baseband in-phase and quadrature components of the drive, respectively. They are defined as:
\begin{equation}
\tilde{I}_\mathrm{in}(t)=V_\mathrm{in}\tilde{p}(t)\,,\quad\tilde{Q}_\mathrm{in}(t)=V_\mathrm{in}\tilde{q}(t)\,.\label{tildeIinQin}
\end{equation}
The power of $s(t)$ that would be dissipated across a 50~resistor in units of dBm is:
\begin{equation}
P_\mathrm{in}=10\,\mathrm{log}_{10}\left(\frac{\langle s^2(t)\rangle}{50\times10^{-3}}\right)=10\,\mathrm{log}_{10}\left(\frac{V_\mathrm{in}^2\langle x^2(t)\rangle}{50\times10^{-3}}\right)\,,
\end{equation}
where $\langle\cdot\rangle$ averages over time. Hence, the peak voltage $V_\mathrm{in}$ reads:
\begin{equation}
V_\mathrm{in}=\left(\frac{10^{P_\mathrm{in}/10}\times50\times10^{-3}}{\langle x^2(t)\rangle}\right)^{1/2}\,.
\end{equation}
The total voltage between the gate and the resonator is the total voltage of the standing wave at the gate:
\begin{equation}
V_\mathrm{d}=(1+\Gamma)V_\mathrm{in}\,,
\end{equation}
where $\Gamma\simeq1$ is the reflection coefficient of $s(t)$ at the gate. The equivalent driving power that would be dissipated across a 50~Ohm resistor in units of dBm is then:
\begin{equation}
P_\mathrm{d}=10\,\mathrm{log}_{10}\left[\frac{(1+\Gamma)^2\langle s^2(t)\rangle}{50\times10^{-3}}\right]=20\,\mathrm{log}_{10}(1+\Gamma)+P_\mathrm{in}\simeq6+P_\mathrm{in}\,.\label{Pd}
\end{equation}
The driving force waveform reads:
\begin{equation}
F_\mathrm{d}(t)=\vert V_\mathrm{g}^\mathrm{dc}\vert(1+\Gamma)\Big\vert\frac{\mathrm{d}C_\mathrm{g}}{\mathrm{d}z}\Big\vert s(t)\,,\label{bigFd}
\end{equation}
where $V_\mathrm{g}^\mathrm{dc}$ is the dc voltage applied between the resonator and the gate, and $\mathrm{d}C_\mathrm{g}/\mathrm{d}z$ is the derivative with respect to the dispacement of the resonator of the capacitance between the resonator and the gate.

\begin{figure}[H]
\centering
\includegraphics{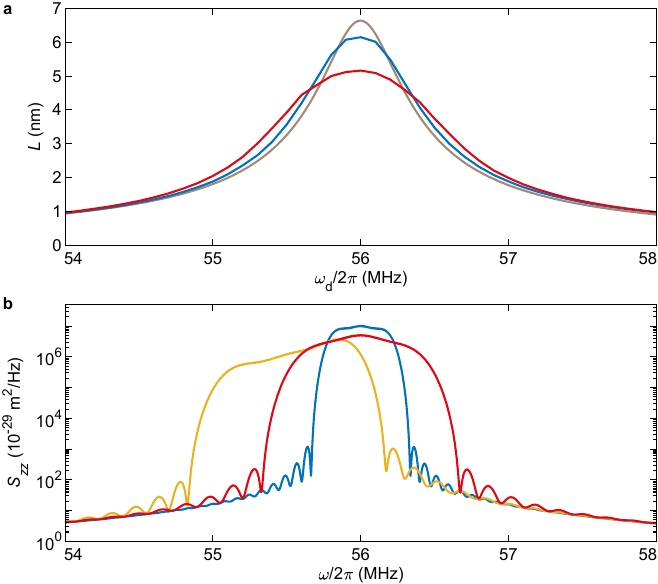}
\caption{Linear response to the multifrequency driving force. (a) $I-Q$ vector length $L$ as a function of angular drive frequency $\omega_\mathrm{d}$ for symbol rate $R_\mathrm{sym}=0.5\times10^6$~Hz (blue trace) and $R_\mathrm{sym}=10^6$~Hz (red trace). The grey trace shows the response to a driving force whose in-phase and quadrature components are not amplitude-modulated, see Eq.~(\ref{unmod}). Driving power $P_\mathrm{d}=-34$~dBm. (b) Averaged single-sided power spectral density of amplitude fluctuations $S_{zz}$ as a function of angular spectral frequency $\omega$ in the linear regime for $P_\mathrm{d}=-31$~dBm. Blue trace: $R_\mathrm{sym}=0.5\times10^6$~Hz, $\omega_\mathrm{d}/2\pi=56\times10^6$~Hz. Red trace: $R_\mathrm{sym}=10^6$~Hz, $\omega_\mathrm{d}/2\pi=56\times10^6$~Hz. Orange trace: $R_\mathrm{sym}=10^6$~Hz, $\omega_\mathrm{d}/2\pi=55.5\times10^6$~Hz. In both panels, driving force waveforms are computed using a raised cosine filter. Quality factor $Q_\mathrm{m}=100$, effective mass $2\times10^{-17}$~kg, resonant frequency 56~MHz.}\label{figlin}
\end{figure}

For comparison, a driving force with the same peak amplitude but whose in-phase and quadrature components are not amplitude-modulated reads:
\begin{equation}
F_\mathrm{d}^\mathrm{unmod}(t)=F_\mathrm{d}^\mathrm{unmod}\left[\cos(\omega_\mathrm{d}t)+\sin(\omega_\mathrm{d}t)\right]=2^{1/2}F_\mathrm{d}^\mathrm{unmod}\cos(\omega_\mathrm{d}t-\pi/4)\,,\label{unmod}
\end{equation}
with
\begin{equation}
F_\mathrm{d}^\mathrm{unmod}=\vert V_\mathrm{g}^\mathrm{dc}\vert(1+\Gamma)V_\mathrm{in}\Big\vert\frac{\mathrm{d}C_\mathrm{g}}{\mathrm{d}z}\Big\vert\,. \label{forcelineaire}
\end{equation}

Supplementary Fig.~\ref{figlin}a shows our calculations of the linear response $L=\langle I_\mathrm{out}+\mathrm{i}Q_\mathrm{out}\rangle$ to $F_\mathrm{d}(t)$ for $P_\mathrm{d}=-34$~dBm, quality factor $Q_\mathrm{m}=100$, effective mass $m_\mathrm{eff}\simeq2\times10^{-17}$~kg, and resonant frequency $\omega_\mathrm{m}/2\pi=56$~MHz. The symbol rates $R_\mathrm{sym}=0.5\times10^6$~Hz (blue trace) and $R_\mathrm{sym}=10^6$~Hz (red trace) are used. Also shown is the response to $F_\mathrm{d}^\mathrm{unmod}(t)$, the unmodulated drive (grey trace). The linewidth broadening of $L(\omega_\mathrm{d})$ as $R_\mathrm{sym}$ increases underlies the fact that, even where $\omega_\mathrm{d}$ is far away from resonance, vibrations can still be driven by the edge of the spectrum of the drive near $\omega_\mathrm{d}\pm\pi W$, where $2\pi W$ is the full bandwidth of the drive. For completeness, Supplementary Fig.~\ref{figlin}b displays the (averaged) single-sided power spectral density of amplitude fluctuations $S_{zz}$ as a function of angular spectral frequency $\omega$ in the linear regime for $P_\mathrm{d}=-31$~dBm. $S_{zz}(\omega)$ reads:
\begin{equation}
S_{zz}(\omega)=\vert\chi(\omega)\vert^2S_{FF}(\omega)\,,\quad\chi(\omega)=\left[m_\mathrm{eff}(\omega_\mathrm{m}^2-\omega^2-\mathrm{i}\omega\omega_\mathrm{m}/Q_\mathrm{m})\right]^{-1}\,,
\end{equation}
where $S_{FF}(\omega)$ is the single-sided power spectral density of $F_\mathrm{d}(t)$ defined as:
\begin{equation}
S_{FF}(\omega)=\frac{2}{t_\mathrm{meas}}\vert\hat{F}_\mathrm{d}(\omega)\vert^2\,,\quad\hat{F}_\mathrm{d}(\omega)=\int_{-\infty}^\infty\mathrm{d}t\,e^{\mathrm{i}\omega t}\,F_\mathrm{d}(t)\,,
\end{equation}
with $t_\mathrm{meas}$ the measurement time of $F_\mathrm{d}(t)$. Supplementary Fig.~\ref{figlin}b shows $S_{zz}(\omega)$ calculated for $R_\mathrm{sym}=0.5\times10^6$~Hz and $\omega_\mathrm{d}=\omega_\mathrm{m}$ (blue trace); $R_\mathrm{sym}=10^6$~Hz and $\omega_\mathrm{d}=\omega_\mathrm{m}$ (red trace); and $R_\mathrm{sym}=10^6$~Hz and $\omega_\mathrm{d}=\omega_\mathrm{m}-\pi R_\mathrm{sym}$ (orange trace).

\subsection{GNU Radio flowgraphs used in this work}

\subsubsection{Preparing and converting video files}
We encode Transport Streams from MP4 videos. To this end, we first identify the transport stream (TS) rate at which the stream will be encoded; it depends on the type of modulation (here, QPSK) and on the code rate. We identify the TS rate using dvbs2rate.c, an open-source utility created by Ron Economos \cite{ron0,Ettus}. We can then convert an MP4 video into a TS file using FFmpeg, an open-source software for processing multimedia files \cite{FFmpeg}. The command line is
\begin{quote}
\textsf{ffmpeg -i video.mp4 -c:v copy -c:a copy -muxrate ts\_rate -f mpegts video.ts}
\end{quote}
where \textsf{video.mp4} is the input video, \textsf{ts\_rate} is the calculated TS rate, and \textsf{video.ts} is the converted video.

At the output of the receiver, we convert $I-Q$ data into a TS file using leandvb \cite{leandvb}, an open-source software for DVB-S and DVB-S2 demodulation (we use GNU Radio for DVB-S and DVBS-2 demodulation and use leandvb for $I-Q$ to TS conversion). The command line is
\begin{quote}
\textsf{./leandvb -{}-gui -v -d -f 1000e3 -{}-sr 1000e3 -{}-cr 5/6 $<$received\_video.iq$>$ received\_video.ts}
\end{quote}
where \textsf{-{}-gui} displays the symbol constellation and the spectrum of the signal, \textsf{-v} and \textsf{-d} output debugging information, \textsf{-f} is the input sampling rate, \textsf{-{}-sr} is the symbol rate (here, $\mathrm{sr}=10^6$~Hz), \textsf{-{}-cr} is the code rate, \textsf{$<$received\_video.iq$>$} is the $I-Q$ stream and \textsf{received\_video.ts} is the output TS. We convert TS into MP4 with FFmpeg using this command line:
\begin{quote}
\textsf{ffmpeg -i received\_video.ts -c:v libx264 received\_video.mp4 }
\end{quote}
For RTL-SDR type receivers such as NESDR SMArt, the maximum sampling rate is $2.4\times10^6$~Hz. Because the bit generation rate is twice the symbol rate for 4-QAM and QPSK (each symbol is defined by two bits), we set the sampling rate of NESDR SMArt to be twice the symbol rate $R_\mathrm{sym}$. That is, we set the sampling rate of the receiver either to 1~MHz ($R_\mathrm{sym}=0.5\times10^6$~Hz) or to 2~MHz ($R_\mathrm{sym}=10^6$~Hz).

\begin{figure}[h]
\centering
\includegraphics[width=16.5cm]{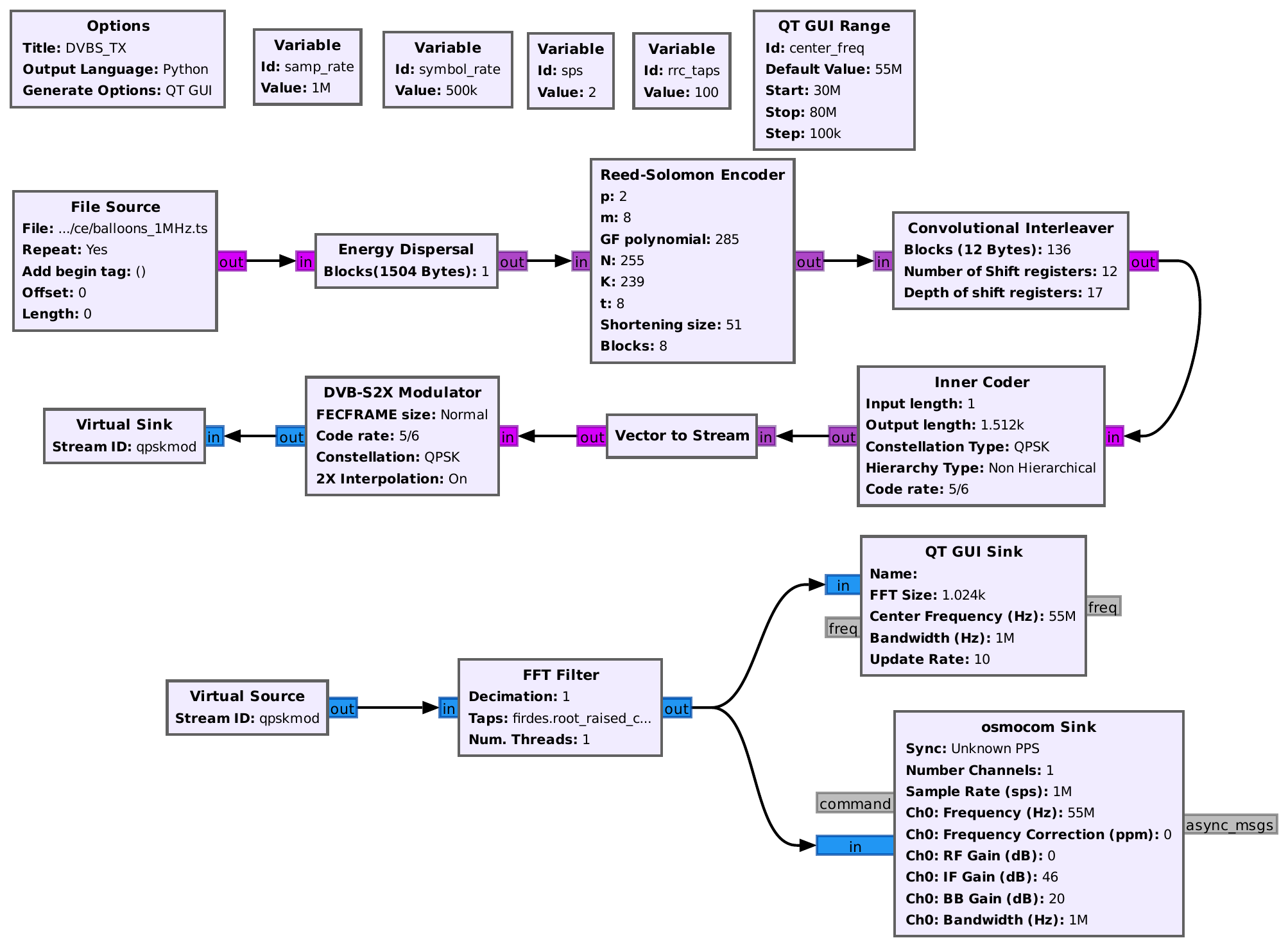}
\caption{GNU Radio flowgraph used to transmit a DVB-S2 video stream.}\label{DVBS2transmit}
\end{figure}

\subsubsection{Transmitting a DVB-S2 video stream}

Supplementary Fig.~\ref{DVBS2transmit} displays the GNU Radio flowgraph used to transmit a DVB-S2 (Digital Video Broadcasting - Satellite - Second Generation) video stream. It is based on GNU Radio modules written by Ron Economos \cite{Ettus,ron1,ron2}. The flowgraph works as follows. \textsf{File Source} reads the Transport Stream (TS) video file and outputs the data stream. \textsf{Energy Dispersal} scrambles the data stream to increase the randomness of the spectrum. \textsf{Reed-Solomon Encoder} encodes the input data block, adding the appropriate number of redundant symbols to achieve forward error correction. \textsf{Convolutional Interleaver} interleaves the output of the Reed-Solomon encoder. The interleaving process allows neighboring data to be dispersed to different locations, so that when an error occurs in the transmission of data, the error is dispersed to different locations without successively affecting the neighboring data. This way of dispersing errors helps to improve the stability of data transmission and reduces the scope of influence of error transmission. We set the parameter \textsf{Constellation Type} in \textsf{Inner Coder} to QPSK; the binary data stream will be converted to the corresponding symbols $\{S_1=0,S_2=1,S_3=2,S_4=3\}$. \textsf{Vector to Stream} converts the vector data stream output by \textsf{Inner Coder} into a continuous data stream. \textsf{DVB-S2X Modulator} maps the input symbols $\{S_1=0,S_2=1,S_3=2,S_4=3\}$ to $(1,\mathrm{i})$, $(-1,\mathrm{i})$, $(-1,-\mathrm{i})$, and $(1,-\mathrm{i})$. \textsf{FFT Filter} and \textsf{DVB-S2X Modulator} are connected via \textsf{Virtual Source} and \textsf{Virtual Sink} modules. \textsf{FFT Filter} uses a root raised cosine filter (RRC) to filter the QPSK signal, thereby minimizing intersymbol interference. \textsf{Taps} is \textsf{firdes.root\_raised\_cosine(1.0,  samp\_rate, samp\_rate/2, 0.35, rrc\_taps)} where \textsf{samp\_rate} is the sample frequency of RRC, \textsf{samp\_rate/2} is the bandwidth of RRC, \textsf{0.35} is the roll-off factor of RRC, and \textsf{rrc\_taps} is 100 and represents the length of the filter. \textsf{osmocom Sink} sets the parameters for HackRF~One, including center frequency (\textsf{Frequency}), \textsf{Bandwidth}, \textsf{RF Gain}, \textsf{IF Gain}, and \textsf{BB Gain} (baseband). \textsf{QT GUI Sink} is used to display the spectrum, the waveform and the constellation diagram of the RRC filtered signal. Note that \textsf{symbol\_rate}, the symbol rate, equates to \textsf{samp\_rate/2}; it is also what we call $R_\mathrm{sym}$ in the main text.

\begin{figure}[t]
\centering
\includegraphics[width=16.5cm]{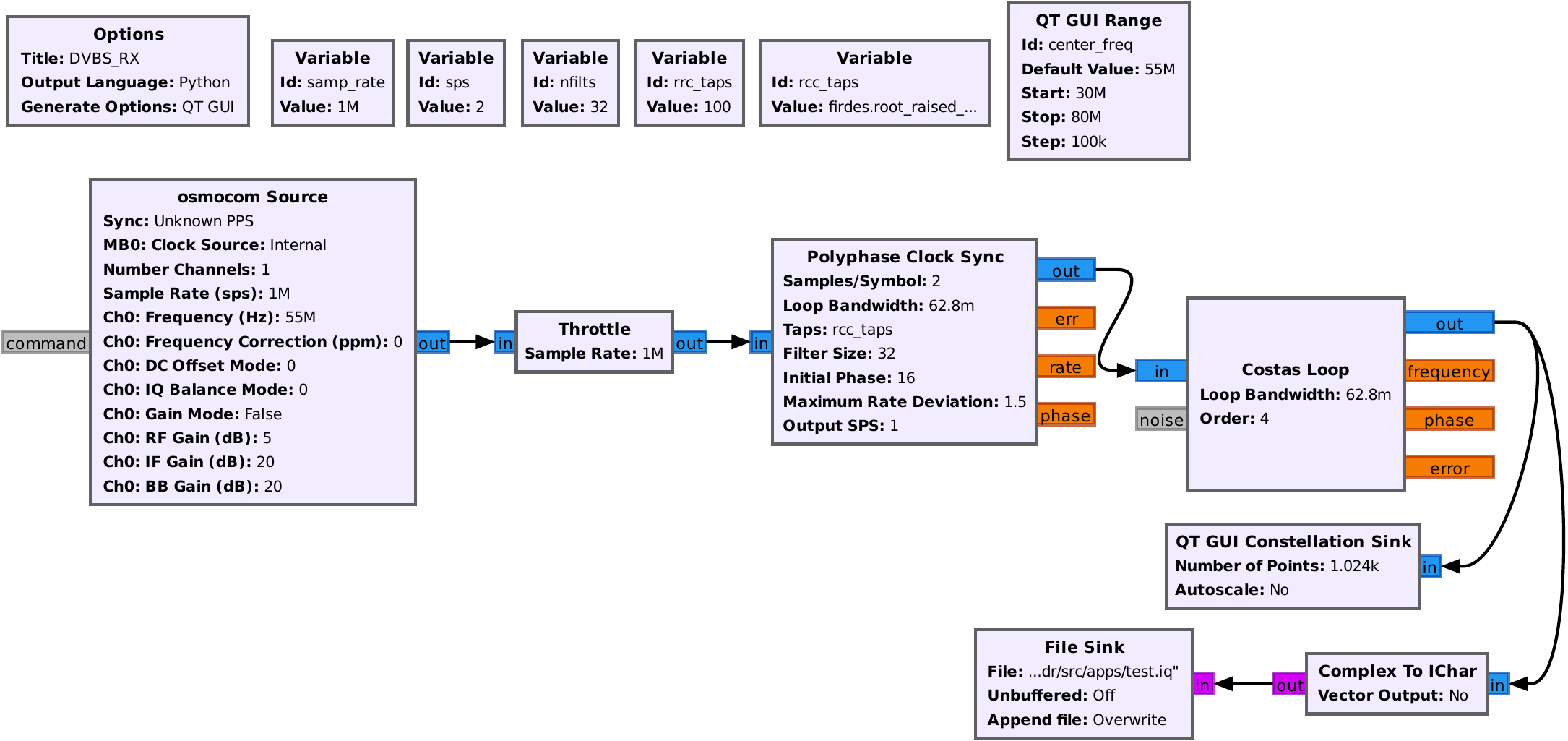}
\caption{GNU Radio flowgraph used to receive a DVB-S2 video stream.}\label{DVBS2receive}
\end{figure}

\subsubsection{Receiving a DVB-S2 video stream}
Supplementary Fig.~\ref{DVBS2receive} displays the GNU Radio flowgraph used to receive a DVB-S2 video stream. \textsf{Osmocom Source} sets parameters for NESDR SMArt (our software-defined radio receiver), including center frequency (\textsf{Frequency}), bandwidth (\textsf{Sample Rate}), \textsf{RF Gain}, \textsf{IF Gain}, and \textsf{BB Gain} (baseband). The center frequency and the bandwidth of NESDR SMArt and those of HackRF~One have to be the same. \textsf{Throttle} is used to limit the rate of data flow to prevent data from being fed into subsequent modules too quickly, causing problems with overloading the system or data loss. \textsf{Polyphase Clock Sync} module implements clock synchronization through a polyphase filter structure, which effectively filters the signal and compensates for clock offsets, thus improving the quality and accuracy of the received signal and minimizing intersymbol interference \cite{Franks1980,Harris2001}. \textsf{Costas Loop} achieves phase synchronization by means of the Costas loop structure, which can effectively phase adjust the signal and compensate for the phase offset, thus improving the phase accuracy and stability of the received signal. \textsf{Complex To IChar} converts stream of complex to a stream of interleaved Chars (a Char is a sequence of bytes that stores one symbol or a single letter). The output stream contains Chars with twice as many output items as input items. \textsf{File Sink} saves the $I-Q$ data (in-phase and quadrature components of the output). \textsf{QT GUI Constellation Sink} displays the $I-Q$ constellation diagram of the baseband of the received signal.

\subsubsection{Measuring the bit error ratio (BER)}
\begin{figure}[h]
\centering
\includegraphics[width=16.5cm]{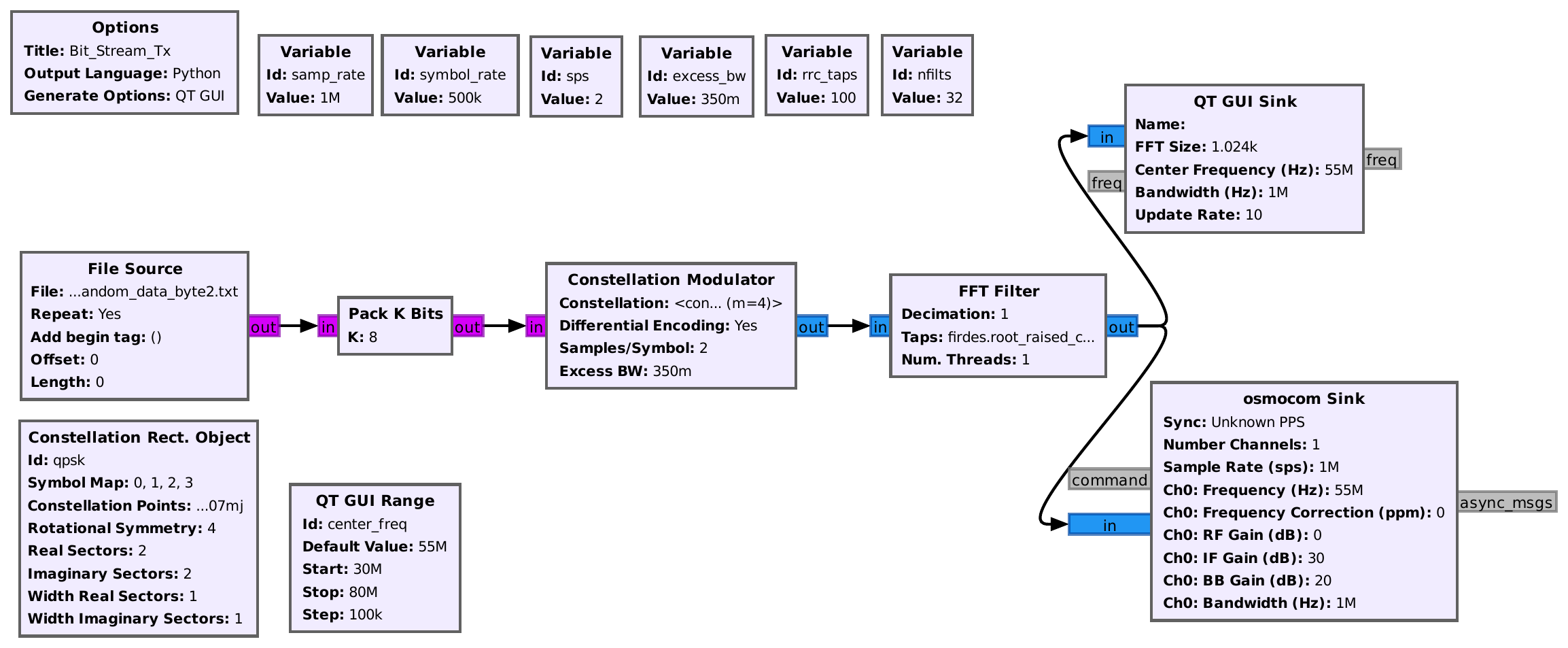}
\caption{GNU Radio flowgraph used to transmit a bit stream for BER measurements.}\label{BERtx}
\end{figure}

Unlike the flowgraph for sending DVB-S2 video streams, here we do not use any error correction algorithm. Supplementary Fig.~\ref{BERtx} displays the GNU Radio flowgraph for sending a bit stream used to measure BER. \textsf{File Source} is used to read the bit stream from a file we prepared (it starts with a visible marker, see Fig.~4a in the main text). \textsf{Pack K Bits} packs the incoming bit stream into one byte for every 8 bits. \textsf{Constellation Modulator} performs QPSK modulation of the incoming byte stream and differentially encodes the symbols. A differential encoder makes transmitted data immune from possible inversions of the binary input waveform. The polarity of a differentially encoded signal can be inverted without affecting the decoded signal. This is important in our case because polarity inversion does take place and is caused by the phase response of the resonator as a function of drive frequency. Next, \textsf{FFT Filter} uses a root raised cosine filter (RRC) to filter the QPSK signal. \textsf{osmocom Sink} sets the parameters for HackRF~One.

\begin{figure}[H]
\centering
\includegraphics[width=16.5cm]{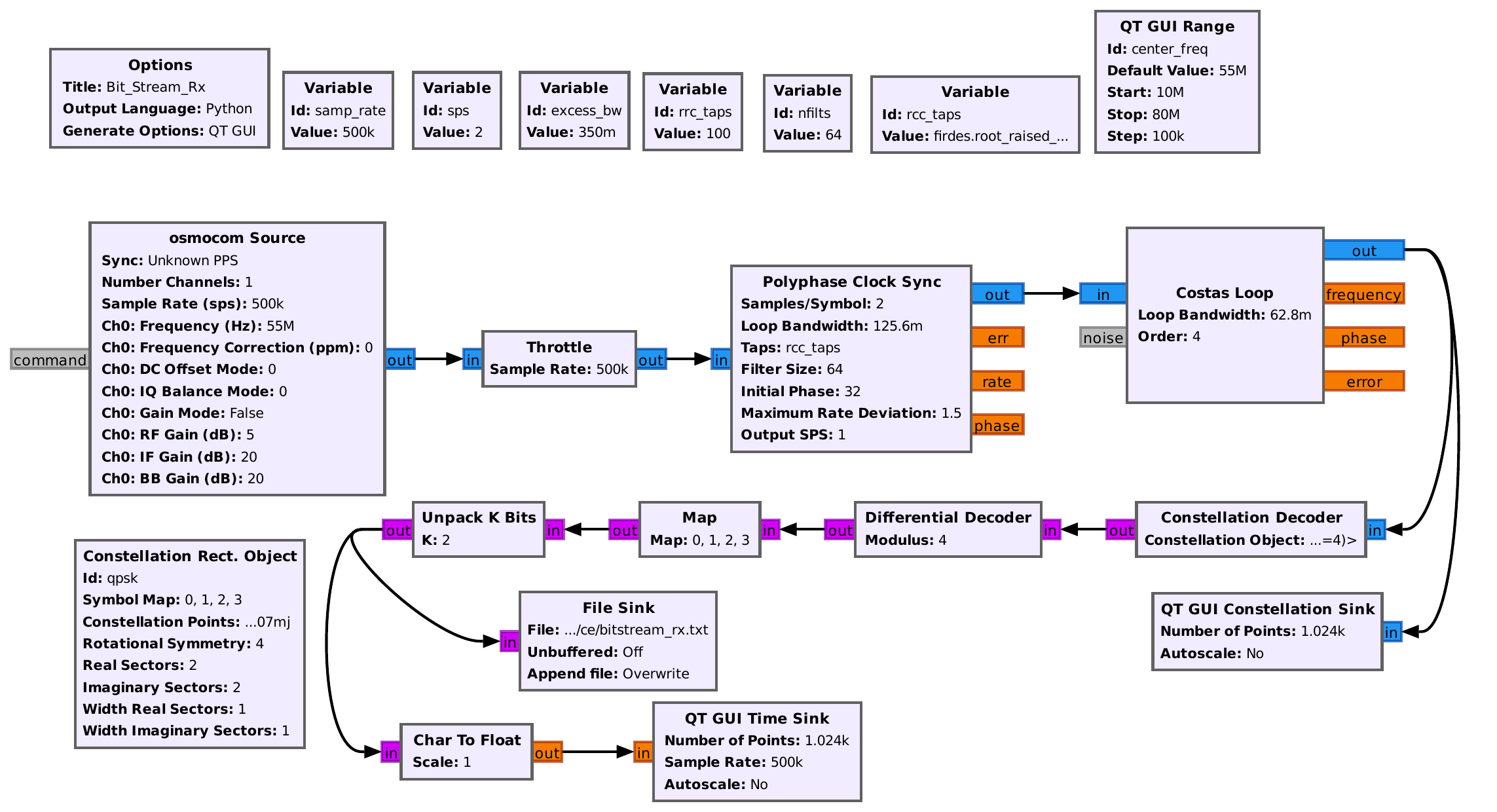}
\caption{GNU Radio flowgraph used to receive a bit stream for BER measurements.}\label{BERrx}
\end{figure}

Supplementary Fig.~\ref{BERrx} displays the GNU Radio flowgraph for receiving a bit stream to measure BER. It works as follows. \textsf{osmocom Source}, \textsf{Polyphase Clock Sync}, \textsf{Costas Loop} and \textsf{QT GUI Contellation Sink} have the same role as in Supplementary Fig.~\ref{DVBS2receive}. After receiving the QPSK signal, \textsf{Constellation Decoder} converts the complex signal into symbols in the QPSK constellation diagram. Using these symbols, \textsf{Differential Decoder} performs a differential demodulation (see above for the role of the differential encoder). \textsf{Modulus} is set to 4, which means that the module will perform QPSK differential demodulation. Next, \textsf{Map} maps the QPSK symbols to $\{0, 1, 2, 3\}$. With \textsf{K} set to 2, \textsf{Unpack K Bits} converts the received digits $\{0, 1, 2, 3\}$ into a 2-bit output. Finally, \textsf{File Sink} saves the received  bit stream. We use \textsf{Char To Float} and \textsf{QT GUI Time Sink} to display the bit stream.

\subsection{$I-Q$ vector length and reciprocal of error vector magnitude at low drive power}

The averaged length of the $I-Q$ vector in one quadrant of the constellation diagram, $L=\langle\vert I_\mathrm{out}+\mathrm{i}Q_\mathrm{out}\vert\rangle$, and the reciprocal of the error vector magnitude $\varepsilon$, both at the output of the photodetector, are shown in Supplementary Fig.~\ref{lowpower0p5} as a function of drive frequency $f_\mathrm{d}$ at a symbol rate $R_\mathrm{sym}=0.5\times10^6$~Hz. Low values for the drive power $P_\mathrm{d}$ are used (see Eq.~(\ref{Pd}) for the definition of $P_\mathrm{d}$). We find that the smallest $\varepsilon$ at which a video signal can be decoded is about 12~dB (peak value of $\varepsilon(f_\mathrm{d})$ in Supplementary Fig.~\ref{lowpower0p5}d).
\begin{figure}[h]
\centering
\includegraphics{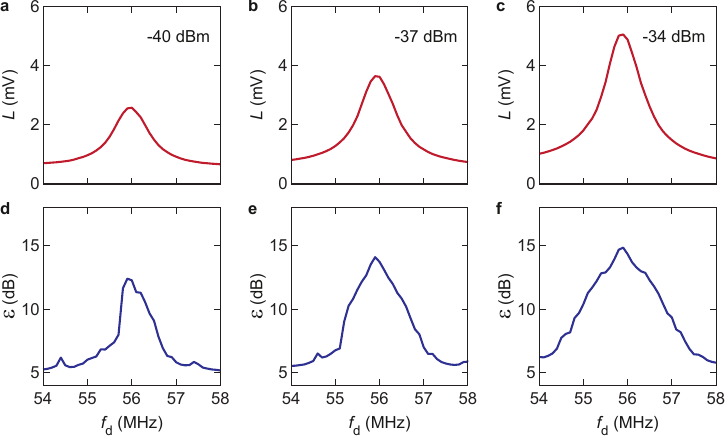}
\caption{$L$ and $\varepsilon$ at low drive power. (a-c) Averaged $I-Q$ vector $L$ and (d-f) reciprocal of error vector magnitude $\varepsilon$ as a function of drive frequency $f_\mathrm{d}$ at low drive power $P_\mathrm{d}$ on the gate and at a symbol rate $R_\mathrm{sym}=0.5\times10^6$~Hz. (a, d) $P_\mathrm{d}=-40$~dBm. (b, e) $P_\mathrm{d}=-37$~dBm. (c, f) $P_\mathrm{d}=-34$~dBm.}\label{lowpower0p5}
\end{figure}

To understand this lower bound, we express $L$ and the characteristic size $d$ of the symbol cloud in one quadrant. $L^2$ reads:
\begin{equation}
L^2=\left(\frac{1}{N}\sum_n I_n\right)^2 + \left(\frac{1}{N}\sum_n Q_n\right)^2=\mu_I^2+\mu_Q^2\,,\label{Lexplicit}
\end{equation}
where $N$ is the number of $(I_n,Q_n)$ measurements in one quadrant. $d^2$ reads:
\begin{equation}
d^2=\frac{1}{N}\sum_n\left[\left(I_n-\mu_I\right)^2+\left(Q_n-\mu_Q\right)^2\right]=\sigma_I^2+\sigma_Q^2\,, \label{dsqexplicit}
\end{equation}
where $\sigma_I^2$ is the variance of $\{I_n\}$ and $\sigma_Q^2$ is the variance of $\{Q_n\}$. At low $P_\mathrm{d}$, the mechanical response is linear. On resonance, $d$ is determined by noise, which we assume to be white with a Gaussian statistics and with a mean $\mu$ and a variance $\sigma^2$. Hence, $\mu_I\simeq\mu_Q\equiv\mu$ and $\sigma_I\simeq\sigma_Q\equiv\sigma$. Using Eqs.~(\ref{Lexplicit}) and (\ref{dsqexplicit})$, \varepsilon$ reads:
\begin{equation}
\varepsilon=10\,\mathrm{log}_{10}\left(\frac{L^2}{d^2}\right)\simeq20\,\mathrm{log}_{10}\left(\frac{\mu}{\sigma}\right)\,.
\end{equation}
In Supplementary Fig.~\ref{varminimum}, we plot a calculated constellation diagram for $\mu/\sigma=4$, which corresponds to $\varepsilon\simeq12$~dB. The symbol clouds are still well separated, each one residing in its own quadrant. Upon reducing $\mu/\sigma$, hence lowering $\varepsilon$, the symbol clouds spread to the other quadrants and the decoding fails.

\begin{figure}[H]
\centering
\includegraphics{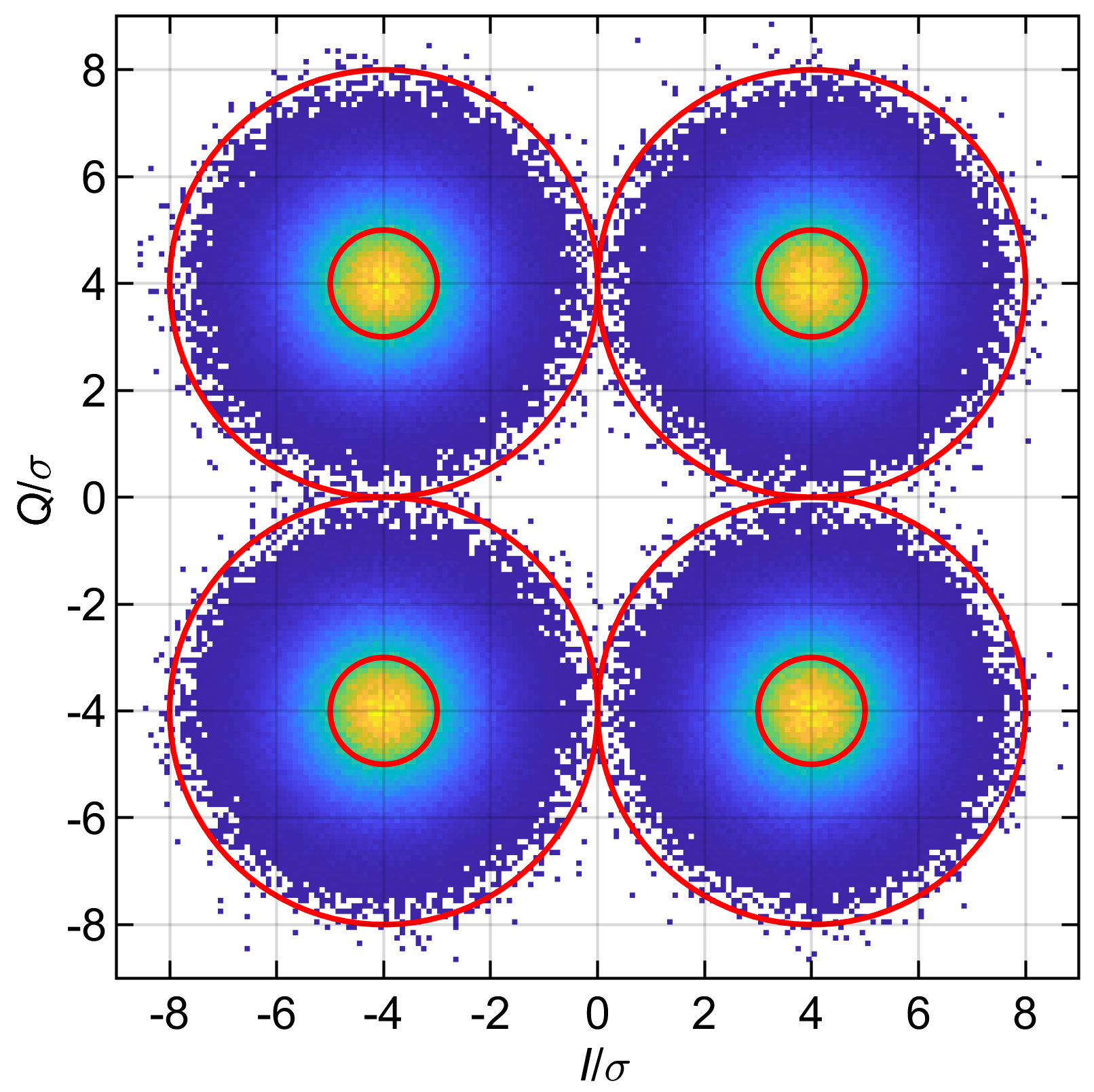}
\caption{Calculated constellation diagram in the case where the spread of the symbol clouds is determined by white Gaussian noise and $\mu/\sigma=4$ (see text for details). Inner circles have a radius of $\sigma$, outer circles have a radius of $4\sigma$.}\label{varminimum}
\end{figure}

\subsection{Estimating the coefficients of the conservative and dissipative forces}\label{coefficients}
\begin{figure}[t]
\centering
\includegraphics{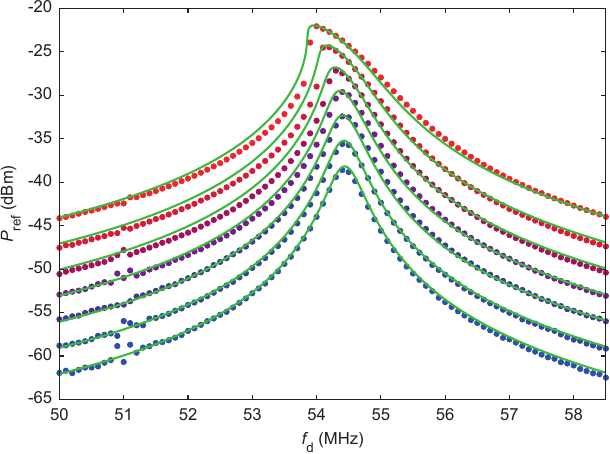}
\caption{Power of the signal at the output of the photodetector $P_\mathrm{ref}$ as a function of drive frequency $f_\mathrm{d}$. Dots: measured data. Lines: fits to Eq.~(\ref{nld}). Drive power at the gate, from bottom to top: $P_\mathrm{single}=-40$, $-37$, $-34$, $-31$, $-28$, $-25$ and $-22$~dBm. Gate voltage $V_\mathrm{g}^\mathrm{dc}=13$~V.}\label{fdPd}
\end{figure}

The coefficients of the conservative and dissipative forces in the equation of motion for the resonator are estimated by measuring the response of the resonator to a single tone driving force. The reponse is measured as an electrical power $P_\mathrm{ref}\propto z^2$ at the output of the photodetector (the index `$\mathrm{ref}$' indicates that the output signal is related to the measured reflected light intensity), where $z$ is the amplitude of vibrations. $P_\mathrm{ref}$ is shown in Supplementary Fig.~\ref{fdPd} as a function of drive frequency $f_\mathrm{d}$ (dots) for various values of the single tone driving power $P_\mathrm{single}$ (see below for the definition of $P_\mathrm{single}$). The response is fit to the solution of the following equation of motion \cite{Dykman1975,Lifshitz2008,Zaitsev2012}:
\begin{equation}
\ddot{z}(t) + \left[\frac{\omega_\mathrm{m}}{Q_\mathrm{m}}+\frac{\eta}{m_\mathrm{eff}}z^2(t)\right]\dot{z}(t) + \left[\omega_\mathrm{m}^2+\frac{\alpha}{m_\mathrm{eff}}z^2(t)\right]z(t)=\frac{F_\mathrm{single}\cos\omega_\mathrm{d}t}{m_\mathrm{eff}}\,,\label{nld}
\end{equation}
where $\omega_\mathrm{m}=2\pi f_\mathrm{m}$ is the angular resonant frequency of the mode, $Q_\mathrm{m}$ is the quality factor of the linear response, $\eta$ is the nonlinear damping coefficient, $\alpha$ is the Duffing elastic constant, $m_\mathrm{eff}$ is the effective mass of the mode, and $F_\mathrm{single}$ is the peak amplitude of the single tone driving force. From the size of the suspended flake measured by atomic force microscopy and from the number of graphene layers estimated from optical reflectometry \cite{Lu2022}, we estimate $m_\mathrm{eff}\simeq2\times10^{-17}$~kg. The peak amplitude of the force reads:
\begin{equation}
F_\mathrm{single}=\vert C_\mathrm{g}^\prime V_\mathrm{g}^\mathrm{dc}(1+\Gamma)V_\mathrm{single}\vert\,,\label{Fd}
\end{equation}
where $C_\mathrm{g}^\prime=\mathrm{d}C_\mathrm{g}/\mathrm{d}z\simeq9\times10^{-10}$~F~m$^{-1}$ is the derivative with respect to flexural displacement of the capacitance between the resonator and the gate, $V_\mathrm{g}^\mathrm{dc}$ is the dc voltage between the resonator and the gate, $\Gamma\simeq1$ is the reflection coefficient at the gate, and $V_\mathrm{single}$ is the peak amplitude of the travelling voltage waveform. Correspondingly, the averaged driving power that would be dissipated across a 50~Ohm resistor is:
\begin{equation}
P_\mathrm{single}=10\,\mathrm{log}_{10}\left[\frac{(1+\Gamma)^2V_\mathrm{single}^2}{2\times50\times10^{-3}}\right]\,,
\end{equation}
in units of dBm. Fitting the measured data to Eq.~(\ref{nld}) [see also Supplementary Information in Ref.~\cite{Lu2022}], we estimate $Q_\mathrm{m}\simeq100$, $\eta\simeq1.2\times10^5$~kg~m$^{-2}$~s$^{-1}$, and $\alpha\simeq-10^{14}$~kg~m$^{-2}$~s$^{-2}$.

\subsection{Noise at the output of the receiver}

\begin{figure}[h]
\centering
\includegraphics{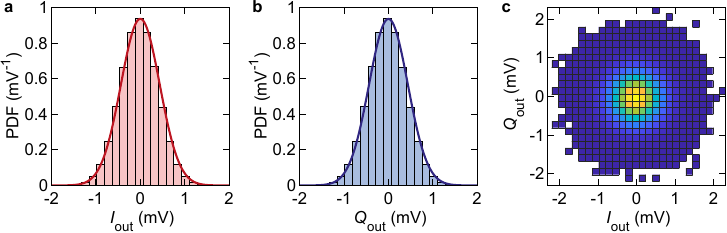}
\caption{Noise at the output of the receiver. (a) Power density function (PDF) of $I_\mathrm{out}$. (b) PDF of $Q_\mathrm{out}$. (c) Distribution of $(I_\mathrm{out},Q_\mathrm{out})$. Traces in (a) and (b) are Gaussian curve fits using the measured variance of $I_\mathrm{out}$ and $Q_\mathrm{out}$. Data in all panels are measured at low drive power $P_\mathrm{d}=-40$~dBm and drive frequency $f_\mathrm{d}=51\times10^6$~Hz far away from resonance.}\label{IQnoise}
\end{figure}

We estimate the noise at the output of the receiver from a large collection of 1.18 million measurements of $(I_\mathrm{out},Q_\mathrm{out})$ at low drive power $P_\mathrm{d}=-40$~dBm and at a drive frequency $f_\mathrm{d}=51\times10^6$~Hz far away from resonance. Supplementary Fig.~\ref{IQnoise}a shows the power density function (PDF) of $I_\mathrm{out}$, Supplementary Fig.~\ref{IQnoise}b shows the PDF of $Q_\mathrm{out}$, and Supplementary Fig.~\ref{IQnoise}c shows the distribution of $(I_\mathrm{out},Q_\mathrm{out})$. From these PDFs and distribution, we conclude that the fluctuations of $I_\mathrm{out}$ and $Q_\mathrm{out}$ follow a white Gaussian statistics. The variance of $I_\mathrm{out}$ and $Q_\mathrm{out}$ is $\sigma^2\simeq0.19$~mV$^2$.

\begin{figure}[t]
\centering
\includegraphics{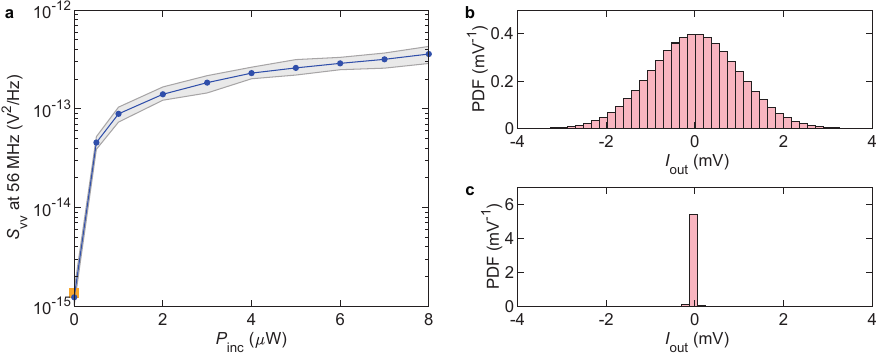}
\caption{Identifying the main source of noise in our measurements. (a) Power spectral density (PSD) of voltage fluctuations at the output of the photodetector at a frequency of 56~MHz. Dots: PSD averaged between 55~MHz and 57~MHz. The upper and lower bounds of the gray shaded area correspond to the maximum and the minimum PSD within this frequency range. (b) Power density function (PDF) of a calibrated white Gaussian noise with a standard deviation of 1~mV measured at the output of the software-defined radio receiver. (c) PDF of the signal at the output of the SDR with the calibration noise source switched off.}\label{APDnoise}
\end{figure}

To narrow down the source of this noise, we separately measure the power spectral density of voltage fluctuations (PSD) at the output of the photodetector and the noise at the output of the software-defined radio receiver. Supplementary Fig.~\ref{APDnoise} shows the PSD at the output of the photodetector measured with a spectrum analyzer at a frequency of 56~MHz (near the mechanical resonant frequency) as a function of optical power $P_\mathrm{inc}$ directly incident on the photodetector. The orange square shows the noise equivalent power of the photodetector specified by the supplier (APD 430A2 by Thorlabs). The dots show PSD averaged between 55~MHz and 57~MHz, and the upper and lower bounds of the gray shaded area correspond to the maximum and the minimum PSD within this frequency range. Given that $P_\mathrm{inc}$ ranges from 2~$\mu$W to 4~$\mu$W in our nanomechanical experiment, and given that our nanomechanical signals are measured within a $\simeq2$~MHz-wide frequency band (the bandwidth of our software-defined radio receiver), we estimate that the variance of voltage fluctuations that originate from the output of the photodetector and that appear at the input of the SDR receiver ranges from $\simeq0.2$~mV$^2$ to $\simeq0.5$~mV$^2$. Our value of $\sigma^2\simeq0.19$~mV$^2$ measured above is at the lower bound of this range. We surmise that digital filtering in our GNU Radio codes, including the root raised cosine filter at the receiver, contributes to lowering the noise we measure in our nanomechanical experiment.

Finally, we estimate the measurement background of our SDR receiver. As an input signal, we use a voltage noise with a Gaussian statistics and a standard deviation of 1~mV within a 100~MHz bandwidth. The variance is calibrated for a 50~Ohm impedance load, which is the input impedance of our SDR receiver. Supplementary Fig.~\ref{APDnoise}b shows the PDF of the in-phase components of the signal at the output of the SDR receiver acquired for 30~s using a receiver bandwidth of 2.4~MHz. Using the same measurement settings, Supplementary Fig.~\ref{APDnoise}c shows the corresponding PDF where the only noise present comes from the 50~Ohm output impedance of the noise source (the output of the source is nominally set to `off'). Using the measurements in Supplementary Fig.~\ref{APDnoise}b as a calibration, we estimate that the smallest amplitude of voltage fluctuations the SDR can measure is larger than 0.1~mV and is limited by digitization error.

Overall, noise in our measurements mostly corresponds to voltage fluctuations at the output of the photodetector. Lowering the intensity of this noise would enable higher order digital modulation ($m$-QAM with $m>4$) at lower drive power.

\subsection{In-phase and quadrature baseband components of the linear response}\label{theory}

\subsubsection{Approach \#1: Homodyne demodulation of the response}

Starting from Eq.~(\ref{tildeIinQin}), we make $\tilde{I}_\mathrm{in}(t)$ and $\tilde{Q}_\mathrm{in}(t)$ periodic with period $\tau$ and we express them as Fourier series:
\begin{eqnarray}
\tilde{I}_\mathrm{in}(t)&=&\sum_{n=0}^{\infty}a_n\cos(n\omega_Lt) + b_n\sin(n\omega_Lt)\\
\tilde{Q}_\mathrm{in}(t)&=&\sum_{n=0}^{\infty}c_n\cos(n\omega_Lt) + d_n\sin(n\omega_Lt)\,,
\end{eqnarray}
with $\omega_\mathrm{L}=2\pi/\tau$, $a_n$, $b_n$ the coefficients of the Fourier series of $\tilde{I}_\mathrm{in}(t)$, and $c_n$, $d_n$ the coefficients of the Fourier series of $\tilde{Q}_\mathrm{in}(t)$. The response to the driving force $F_\mathrm{d}(t)$ defined by Eq.~(\ref{bigFd}) is a displacement $z(t)=F_\mathrm{d}(t)\ast\chi(t)$, where $\chi(t)$ is the linear susceptibility of the resonator and $\ast$ is the convolution operator. The Fourier transform of $\chi(t)$ is $\hat{\chi}(\omega)=[\mathrm{m}_\mathrm{eff}(\omega_\mathrm{m}^2-\omega^2-\mathrm{i}\omega_\mathrm{m}\omega/Q_\mathrm{m})]^{-1}$, where $\omega_\mathrm{m}$ is the resonant angular frequency of the mode, $Q_\mathrm{m}$ is its quality factor, and $\mathrm{m}_\mathrm{eff}$ is its effective mass. At the output of the photodetector, $z$ is linearly converted  into a voltage $v(t)=\xi z(t)$, where $\xi$ is a transduction factor from meter to Volt. This output voltage reads:
\begin{equation}
v(t)=\tilde{I}_\mathrm{out}(t)\cos\omega_\mathrm{d}t+\tilde{Q}_\mathrm{out}(t)\sin\omega_\mathrm{d}t\,,
\end{equation}
where $\tilde{I}_\mathrm{out}(t)$ and $\tilde{Q}_\mathrm{out}(t)$ are the in-phase and quadrature components of the baseband signal, respectively. The latter are obtained by demodulating $v(t)$ as follows:
\begin{eqnarray}
\tilde{I}_\mathrm{out}(t)&=&2h(t)\ast[v(t)\cos\omega_\mathrm{d}t]\\
\tilde{Q}_\mathrm{out}(t)&=&2h(t)\ast[v(t)\sin\omega_\mathrm{d}t]\,,
\end{eqnarray}
where $h$ is a lowpass filter. Writing $\omega_n=n\omega_L$, $\chi=\vert\chi\vert\exp(\mathrm{i}\phi)$, and $\kappa=\vert\xi(1+\Gamma)C_\mathrm{g}^\prime V_\mathrm{g}^\mathrm{dc}\vert$, we find
\begin{equation}
\begin{aligned}
\tilde{I}_\mathrm{out}(t)&=&\frac{\kappa}{4}\sum_{n=1}^{\infty}\vert\chi\vert(\omega_\mathrm{d}+\omega_n)\Bigl\{(a_n-d_n)\cos[\omega_nt-\phi(\omega_\mathrm{d}+\omega_n)]+(b_n+c_n)\sin[\omega_nt-\phi(\omega_\mathrm{d}+\omega_n)]\Bigr\}\\
&+&\frac{\kappa}{4}\sum_{n=1}^{\infty}\vert\chi\vert(\omega_\mathrm{d}-\omega_n)\Bigl\{(a_n+d_n)\cos[\omega_nt+\phi(\omega_\mathrm{d}-\omega_n)]+(b_n-c_n)\sin[\omega_nt+\phi(\omega_\mathrm{d}-\omega_n)]\Bigr\}
\end{aligned}
\end{equation}
and
\begin{equation}
\begin{aligned}
\tilde{Q}_\mathrm{out}(t)&=&\frac{\kappa}{4}\sum_{n=1}^{\infty}\vert\chi\vert(\omega_\mathrm{d}+\omega_n)\Bigl\{(b_n+c_n)\cos[\omega_nt-\phi(\omega_\mathrm{d}+\omega_n)]+(-a_n+d_n)\sin[\omega_nt-\phi(\omega_\mathrm{d}+\omega_n)]\Bigr\}\\
&+&\frac{\kappa}{4}\sum_{n=1}^{\infty}\vert\chi\vert(\omega_\mathrm{d}-\omega_n)\Bigl\{(-b_n+c_n)\cos[\omega_nt+\phi(\omega_\mathrm{d}-\omega_n)]+(a_n+d_n)\sin[\omega_nt+\phi(\omega_\mathrm{d}-\omega_n)]\Bigr\}
\end{aligned}
\end{equation}

For a signal whose full spectral bandwidth $2\pi W$ is smaller than the mechanical linewidth $\omega_\mathrm{m}/Q_\mathrm{m}$, far below resonance $\omega_\mathrm{d}\ll\omega_\mathrm{m}$ the result simplifies to
\begin{eqnarray}
\tilde{I}_\mathrm{out}(t)&\simeq&\frac{\kappa\vert\chi\vert(\omega_\mathrm{d})}{2}\left[\tilde{I}_\mathrm{in}(t)\cos\phi(\omega_\mathrm{d})-\tilde{Q}_\mathrm{in}(t)\sin\phi(\omega_\mathrm{d})\right]\simeq\frac{\kappa\vert\chi\vert(\omega_\mathrm{d})}{2}\tilde{I}_\mathrm{in}(t)\\
\tilde{Q}_\mathrm{out}(t)&\simeq&\frac{\kappa\vert\chi\vert(\omega_\mathrm{d})}{2}\left[\tilde{I}_\mathrm{in}(t)\sin\phi(\omega_\mathrm{d})+\tilde{Q}_\mathrm{in}(t)\cos\phi(\omega_\mathrm{d})\right]\simeq\frac{\kappa\vert\chi\vert(\omega_\mathrm{d})}{2}\tilde{Q}_\mathrm{in}(t)\,.
\end{eqnarray}
Note that both $\tilde{I}_\mathrm{out}(t)$ and $\tilde{Q}_\mathrm{out}(t)$ contain components of $\tilde{I}_\mathrm{in}(t)$ and $\tilde{Q}_\mathrm{in}(t)$. Minimizing either one of these components requires $\omega_\mathrm{d}\ll\omega_\mathrm{m}$ or $\omega_\mathrm{d}\gg\omega_\mathrm{m}$, provided that the measurement noise is weak enough that the off-resonance response can be resolved.

On resonance, $\omega_\mathrm{d}=\omega_\mathrm{m}$, and still assuming $2\pi W<\omega_\mathrm{m}/Q_\mathrm{m}$, the result simplifies to
\begin{eqnarray}
\tilde{I}_\mathrm{out}(t)&\simeq&\frac{\kappa}{2}\sum_{n=1}^{\infty}\vert\chi\vert(\omega_\mathrm{d}+\omega_n)\Bigl\{c_n\sin[\omega_nt-\phi(\omega_\mathrm{d}+\omega_n)]-d_n\cos[\omega_nt-\phi(\omega_\mathrm{d}+\omega_n)]\Bigr\}\nonumber\\
&\simeq&\frac{\kappa\vert\chi\vert(\omega_\mathrm{m})}{2}\tilde{Q}_\mathrm{in}(t)\label{iq1}\\
\tilde{Q}_\mathrm{out}(t)&\simeq&\frac{\kappa}{2}\sum_{n=1}^{\infty}\vert\chi\vert(\omega_\mathrm{d}+\omega_n)\Bigl\{-a_n\sin[\omega_nt-\phi(\omega_\mathrm{d}+\omega_n)]+b_n\cos[\omega_nt-\phi(\omega_\mathrm{d}+\omega_n)]\Bigr\}\nonumber\\
&\simeq&-\frac{\kappa\vert\chi\vert(\omega_\mathrm{m})}{2}\tilde{I}_\mathrm{in}(t)\label{iq2}\,.
\end{eqnarray}
In this case, $\tilde{I}_\mathrm{out}(t)$ only contains components of $\tilde{Q}_\mathrm{in}(t)$, and $\tilde{Q}_\mathrm{out}(t)$ only contains components of $\tilde{I}_\mathrm{in}(t)$. This is a consequence of the symmetry of the linear response at resonance:
\begin{equation}
\vert\chi\vert(\omega_\mathrm{m}+\omega_n)=\vert\chi\vert(\omega_\mathrm{m}-\omega_n)\,,\quad\phi(\omega_\mathrm{m}+\omega_n)=-\phi(\omega_\mathrm{m}-\omega_n)-\pi\,.
\end{equation}

For a signal with $2\pi W\gtrsim\omega_\mathrm{m}/Q_\mathrm{m}$ and $\pi W\simeq\vert \omega_\mathrm{m}-\omega_\mathrm{d}\vert$, below resonance $\omega_\mathrm{d}<\omega_\mathrm{m}$ the result simplifies to
\begin{equation}
\begin{aligned}
\tilde{I}_\mathrm{out}(t)&\simeq&\frac{\kappa}{4}\sum_{n=1}^{\infty}\vert\chi\vert(\omega_\mathrm{d}+\omega_n)\Bigl\{(a_n-d_n)\cos[\omega_nt-\phi(\omega_\mathrm{d}+\omega_n)]+(b_n+c_n)\sin[\omega_nt-\phi(\omega_\mathrm{d}+\omega_n)]\Bigr\}\\
\tilde{Q}_\mathrm{out}(t)&\simeq&\frac{\kappa}{4}\sum_{n=1}^{\infty}\vert\chi\vert(\omega_\mathrm{d}+\omega_n)\Bigl\{(b_n+c_n)\cos[\omega_nt-\phi(\omega_\mathrm{d}+\omega_n)]+(-a_n+d_n)\sin[\omega_nt-\phi(\omega_\mathrm{d}+\omega_n)]\Bigr\}\label{eqoffresonance}
\end{aligned}
\end{equation}
Equation~\ref{eqoffresonance} shows that the in-phase and quadrature components of the demodulated output signal both contain dephased frequency components of $\tilde{I}_\mathrm{in}(t)$ and $\tilde{Q}_\mathrm{in}(t)$. As a result, the decimation of $\tilde{I}_\mathrm{out}(t)$ and $\tilde{Q}_\mathrm{out}(t)$ yields faulty symbols that are unrelated to the original video signal.

\subsubsection{Approach \#2: In-phase and quadrature components of the response in the frame rotating at $\omega_\mathrm{d}$}

In the linear regime, Eq.~(\ref{nld}) simplies to
\begin{equation}
\ddot{z}(t) + 2\Gamma_\mathrm{m}\dot{z}(t)+\omega_\mathrm{m}^2z(t)=\frac{F_\mathrm{d}(t)}{m_\mathrm{eff}}\,,\label{linearregime}
\end{equation}
with $2\Gamma_\mathrm{m}=\omega_\mathrm{m}/Q_\mathrm{m}$. Modulations of $F_\mathrm{d}(t)$ that occur on a time scale much shorter than the characteristic response time of the resonator $1/\Gamma_\mathrm{m}$ are filtered out by the resonator. Modulations on a time scale close to or longer than $1/\Gamma_\mathrm{m}$ lead to slow changes in the amplitude of displacement $u(t)$ on the time scale $1/\Gamma_\mathrm{m}$. To capture this behavior, it is convenient to make the change from the fast oscillating variable $z(t)$ to the slow complex resonator amplitude $u(t)$:
\begin{equation}
%z(t)=u(t)\exp(\mathrm{i}\omega_\mathrm{d}t)+u^*(t)\exp(-\mathrm{i}\omega_\mathrm{d}t)\,,
z(t)=u(t)e^{\mathrm{i}\omega_\mathrm{d}t}+u^*(t)e^{-\mathrm{i}\omega_\mathrm{d}t}\,,
\end{equation}
with $\omega_\mathrm{d}$ the carrier frequency in the drive signal, see Eq.~(\ref{s}). We treat the modulations of $F_\mathrm{d}(t)$ as a perturbation that impacts the response of the resonator only weakly. This means that the velocity $\dot{z}(t)$ should have the same form with and without perturbation \cite{Bogoliubov,Yaxing}. To see
this, consider the harmonic displacement
\begin{equation}
z(t) = a(t)\cos[\omega_\mathrm{d}t+\theta(t)]\,.
\end{equation}
The velocity is
\begin{equation}
\dot{z}(t) = \dot{a}(t)\cos[\omega_\mathrm{d}t+\theta(t)] -a(t)\omega_\mathrm{d}\sin[\omega_\mathrm{d}t+\theta(t)] - a(t)\dot{\theta}(t)\sin[\omega_\mathrm{d}t+\theta(t)]\,.
\end{equation}
In the absence of a perturbation, $a(t)$ and $\theta(t)$ are constant, hence $\dot{z}(t)=-a\omega_\mathrm{d}\sin(\omega_\mathrm{d}t+\theta)$. For $\dot{z}(t)$ to have such a form in the presence of a perturbation, the following should hold:
\begin{equation}
\dot{a}(t)\cos[\omega_\mathrm{d}t+\theta(t)]-a(t)\dot{\theta}(t)\sin[\omega_\mathrm{d}t+\theta(t)]\equiv0\,,\label{vanderpol}
\end{equation}
a condition known as the Van der Pol approximation. Writing $u(t)$ as
\begin{equation}
%u(t)=\frac{a(t)}{2}\exp[\mathrm{i}\theta(t)]\,,\quad u^*(t)=\frac{a(t)}{2}\exp[-\mathrm{i}\theta(t)]\,,
u(t)=\frac{a(t)}{2}e^{\mathrm{i}\theta(t)}\,,\quad u^*(t)=\frac{a(t)}{2}e^{-\mathrm{i}\theta(t)}\,,
\end{equation}
Eq.~(\ref{vanderpol}) leads to
\begin{equation}
%\dot{u}(t)\exp(\mathrm{i}\omega_\mathrm{d}t)+\dot{u}^*(t)\exp(-\mathrm{i}\omega_\mathrm{d}t)=\dot{a}(t)\cos[\omega_\mathrm{d}t+\theta(t)]-a(t)\dot{\theta}(t)\sin[\omega_\mathrm{d}t+\theta(t)]\equiv0\,.
\dot{u}(t)e^{\mathrm{i}\omega_\mathrm{d}t}+\dot{u}^*(t)e^{-\mathrm{i}\omega_\mathrm{d}t}=\dot{a}(t)\cos[\omega_\mathrm{d}t+\theta(t)]-a(t)\dot{\theta}(t)\sin[\omega_\mathrm{d}t+\theta(t)]\equiv0\,.
\end{equation}
Using this expression, we express $\dot{z}(t)$ as
\begin{equation}
%\dot{z}(t)=\mathrm{i}\omega_\mathrm{d}\left[u(t)\exp(\mathrm{i}\omega_\mathrm{d}t) - u^*(t)\exp(-\mathrm{i}\omega_\mathrm{d}t)\right]\,,
\dot{z}(t)=\mathrm{i}\omega_\mathrm{d}\left[u(t)e^{\mathrm{i}\omega_\mathrm{d}t} - u^*(t)e^{-\mathrm{i}\omega_\mathrm{d}t}\right]\,,
\end{equation}
and we express $\ddot{z}(t)$ as
\begin{equation}
%\ddot{z}(t)=2\mathrm{i}\omega_\mathrm{d}\dot{u}(t)\exp(\mathrm{i}\omega_\mathrm{d}t) - \omega_\mathrm{d}^2z(t)\,.
\ddot{z}(t)=2\mathrm{i}\omega_\mathrm{d}\dot{u}(t)e^{\mathrm{i}\omega_\mathrm{d}t} - \omega_\mathrm{d}^2z(t)\,.
\end{equation}
Plugging $z(t)$, $\dot{z}(t)$ ans $\ddot{z}(t)$ into Eq.~(\ref{linearregime}), we obtain the equation of motion for the slow amplitude $u$:
\begin{equation}
%\dot{u}(t)+(\Gamma+\mathrm{i}\Delta\omega)u(t)=f_u(t)=-\mathrm{i}\frac{F_\mathrm{d}}{4m_\mathrm{eff}\omega_\mathrm{d}}\left[I_\mathrm{in}(t)-\mathrm{i}Q_\mathrm{in}(t)\right]\,,
\dot{u}(t)+(\Gamma+\mathrm{i}\Delta\omega)u(t)=-\mathrm{i}\frac{\zeta}{4m_\mathrm{eff}\omega_\mathrm{d}}\left[\tilde{I}_\mathrm{in}(t)-\mathrm{i}\tilde{Q}_\mathrm{in}(t)\right]\,,
\end{equation}
with
\begin{equation}
\Delta\omega=-\frac{\omega_\mathrm{m}^2-\omega_\mathrm{d}^2}{2\omega_\mathrm{d}}\simeq\omega_\mathrm{d}-\omega_\mathrm{m}\,,\label{EQMu1}
\end{equation}
where $\zeta=\vert V_\mathrm{g}^\mathrm{dc}(1+\Gamma)C_\mathrm{g}^\prime\vert$ is defined in Eq.~(\ref{bigFd}), and $\tilde{I}_\mathrm{in}(t)$, $\tilde{Q}_\mathrm{in}(t)$ are defined in Eq.~(\ref{s}). The solution to Eq.~(\ref{EQMu1}) is
\begin{equation}
u(t)=-\mathrm{i}\frac{\zeta}{4m_\mathrm{eff}\omega_\mathrm{d}}\int_{-\infty}^t \mathrm{d}t^\prime\left[\tilde{I}_\mathrm{in}(t^\prime)-\mathrm{i}\tilde{Q}_\mathrm{in}(t^\prime)\right]\exp\left[-(\Gamma_\mathrm{m}+\mathrm{i}\Delta\omega)(t-t^\prime)\right]\,.\label{solu}
\end{equation}
To clarify the interplay of the signal bandwidth and the resonator bandwidth, we express $\tilde{I}_\mathrm{in}(t)$ and $\tilde{Q}_\mathrm{in}(t)$ as their Fourier transforms $\hat{I}_\mathrm{in}(\omega)$ and $\hat{Q}_\mathrm{in}(\omega)$, which we define as:
\begin{equation}
\tilde{I}_\mathrm{in}(t)=\frac{1}{2\pi}\int_{-\infty}^{\infty}\mathrm{d}\omega\,\hat{I}_\mathrm{in}(\omega)e^{-\mathrm{i}\omega t}\,,\quad \tilde{Q}_\mathrm{in}(t)=\frac{1}{2\pi}\int_{-\infty}^{\infty}\mathrm{d}\omega\,\hat{Q}_\mathrm{in}(\omega)e^{-\mathrm{i}\omega t}\,.
\end{equation}

Equation~(\ref{solu}) simplifies to
\begin{eqnarray}
u(t)&=&-\mathrm{i}\frac{\zeta}{8\pi m_\mathrm{eff}\omega_\mathrm{d}}\int_{-\infty}^\infty\mathrm{d}\omega\,\frac{\hat{I}_\mathrm{in}(\omega)-\mathrm{i}\hat{Q}_\mathrm{in}(\omega)}{\Gamma_\mathrm{m}-\mathrm{i}(\omega-\Delta\omega)}\,e^{-\mathrm{i}\omega t}\nonumber\\
&=&-\mathrm{i}\frac{\zeta}{8\pi m_\mathrm{eff}\omega_\mathrm{d}}\int_{-\pi W}^{\pi W}\mathrm{d}\omega\,\frac{\hat{I}_\mathrm{in}(\omega)-\mathrm{i}\hat{Q}_\mathrm{in}(\omega)}{\left[\Gamma_\mathrm{m}^2+(\omega-\Delta\omega)^2\right]^{1/2}}\,e^{\mathrm{i}\phi(\omega)}e^{-\mathrm{i}\omega t}\,,\label{EQMu2}
\end{eqnarray}
with $\phi(\omega)=\mathrm{arctan}[(\omega-\Delta\omega)/\Gamma_\mathrm{m}]$, and where we have integrated over the signal bandwidth $2\pi W$. Based on Eq.~(\ref{EQMu2}), we have $\tilde{I}_\mathrm{out}(t)\propto\mathrm{Re}[u(t)]$ and $\tilde{Q}_\mathrm{out}(t)\propto\mathrm{Im}[u(t)]$. If $2\pi W\ll2\Gamma_\mathrm{m}$, the response of the resonator to the multifrequency drive is only weakly frequency dependent. As a result, the signal-to-noise ratio $\varepsilon$ is large and the bit error ratio BER is low, provided that the measurement noise is low.

If $\Delta\omega=0$, Eq.~(\ref{EQMu2}) reads
\begin{equation}
u(t)=-\mathrm{i}\frac{\zeta}{4\pi m_\mathrm{eff}\omega_\mathrm{d}}\int_{0}^{\pi W}\mathrm{d}\omega\,\frac{\vert\hat{I}_\mathrm{in}(\omega)\vert\cos[\omega t-\phi_I(\omega)-\phi(\omega)]-\mathrm{i}\vert\hat{Q}_\mathrm{in}(\omega)\vert\cos[\omega t-\phi_Q(\omega)-\phi(\omega)]}{(\Gamma_\mathrm{m}^2+\omega^2)^{1/2}}\,,\label{EQMu3}
\end{equation}
where $\phi_I(\omega)$ and $\phi_Q(\omega)$ are the arguments of $\hat{I}_\mathrm{in}(\omega)$ and $\hat{Q}_\mathrm{in}(\omega)$, respectively. We have used $\hat{I}_\mathrm{in}(-\omega)=\hat{I}_\mathrm{in}^*(\omega)$, $\hat{Q}_\mathrm{in}(-\omega)=\hat{Q}_\mathrm{in}^*(\omega)$ since $\tilde{I}_\mathrm{in}(t)$ and $\tilde{Q}_\mathrm{in}(t)$ are real. Equation~(\ref{EQMu3}) shows that $\tilde{I}_\mathrm{out}(t)\propto\tilde{Q}_\mathrm{in}(t)$ and $\tilde{Q}_\mathrm{out}(t)\propto-\tilde{I}_\mathrm{in}(t)$, provided that $2\pi W<2\Gamma_\mathrm{m}$. As a result, $\varepsilon$ is large and BER is low. 

We now consider the magnitude $\vert\chi_u(\omega)\vert=\left[\Gamma^2+(\omega-\Delta\omega)^2\right]^{-1/2}$ and the argument $\phi(\omega)$ of the response in Eq.~(\ref{EQMu2}) in the case where $\Delta\omega=\pi W$ and $2\pi W\simeq2\Gamma_\mathrm{m}$. For $\omega=-\pi W$, $\vert\chi_u\vert=1/(\Gamma_\mathrm{m}\sqrt{5})$ and $\phi\simeq-\pi/2.84$. For $\omega=\pi W$, $\vert\chi_u\vert=1/\Gamma_\mathrm{m}$ and $\phi=0$. This means that the amplitude and the phase of the frequency components of the drive signal are nonuniformly transformed. In particular, $\phi(\omega=0)=-\pi/4$ means that $\tilde{I}_\mathrm{out}(t)$ and $\tilde{Q}_\mathrm{out}(t)$ each contain a superposition of frequency components of both $\tilde{I}_\mathrm{in}(t)$ and $\tilde{Q}_\mathrm{in}(t)$. In other words, $\tilde{I}_\mathrm{out}(t)$ contains frequency components of $\tilde{I}_\mathrm{in}(t)$ and $\tilde{Q}_\mathrm{in}(t)$, and so does $\tilde{Q}_\mathrm{out}(t)$. As a result, $\varepsilon(\omega_\mathrm{d})$ displays a local minimum and $\mathrm{BER}(\omega_\mathrm{d})$ displays a local maximum. 

\subsection{Switching of symbols induced by the phase response near resonance}

In the linear regime and with the drive frequency $f_\mathrm{d}$ set near resonance, Eqs.~(\ref{iq1}) and (\ref{iq2}) yield $I_\mathrm{out,n}=Q_\mathrm{in}/V_\mathrm{in}$ and $Q_\mathrm{out,n}=-I_\mathrm{in}/V_\mathrm{in}$, where $I_\mathrm{out,n}=\mathrm{sgn}(I_\mathrm{out})$ and $Q_\mathrm{out,n}=\mathrm{sgn}(Q_\mathrm{out})$ [see Eq.~(\ref{tildeIinQin}) for $V_\mathrm{in}$]. This represents a switch of symbols induced by the phase response. The following table shows how the in-phase and quadrature components are transformed: input and output symbols are simply shifted by $\pi/2$. This shift is of no consequence to our experimental demodulation, which is based on a differential QPSK encoder that uses the phase difference between adjacent symbols \cite{dQPSK,diffencoder}.

\begin{center}
\begin{tabular}{ |c|c|c| } 
\hline
state & input & output \\
\hline
\multirow{4}{4em}{symbols} & \{1,1\} & \{1,-1\} \\ 
& \{1,-1\}& \{-1,-1\} \\ 
& \{-1,-1\} & \{-1,1\} \\ 
& \{-1,1\} & \{1,1\} \\ 
\hline
\multirow{4}{4em}{bits} & (11) & (10) \\ 
& (10) & (00) \\ 
& (00) & (01) \\ 
& (01) & (11) \\ 
\hline
\end{tabular}
\end{center}

\begin{figure}[t]
\centering
\includegraphics{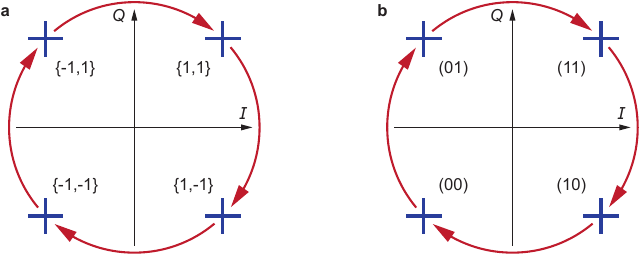}
\caption{Effect of $\pi/2$ dephasing of the response on the output symbols and output bits. The constellation diagram is labelled with symbols in (a) and with pairs of bits in (b). The arrows indicate the transformation from input to output symbols and bits.}\label{constellationBER}
\end{figure}

\newpage

\subsection{Dip in $\varepsilon$ and peak in BER in the nonlinear regime}

The dip in $\varepsilon(f_\mathrm{d})$ and the peak in $\mathrm{BER}(f_\mathrm{d})$ are striking features in our measurements. We address them theoretically in the case of the linear regime in Supplementary Note~\ref{theory}. To understand how they relate to $L(f_\mathrm{d})$ in the nonlinear regime, in Supplementary Fig.~\ref{nonlinearSI} we show our calculations for $L$ [Supplementary Fig.~\ref{nonlinearSI}a], $\varepsilon$ [Supplementary Fig.~\ref{nonlinearSI}b], and BER [Supplementary Fig.~\ref{nonlinearSI}c] as a function of $f_\mathrm{d}$ at $R_\mathrm{sym}=10^6$~Hz, for $P_\mathrm{d}$ ranging from $-31$~dBm to $-4$~dBm, and without addititive white Gaussian noise. The dip in $\varepsilon(f_\mathrm{d})$ and the peak in $\mathrm{BER}(f_\mathrm{d})$ appear concurrently.

The inset to Supplementary Fig.~\ref{nonlinearSI}c shows additional calculations with a finer $f_\mathrm{d}$ step where we resolve the dip in $\varepsilon(f_\mathrm{d})$ and the corresponding $I-Q$ vector length at the dip, $L_\mathrm{dip}$, for $P_\mathrm{d}$ ranging from $-34$~dBm to $0$~dBm. There, the frequency shift $\Delta f_\mathrm{dip}$ of the dip is plotted as a function of the square of the change in vector length $\Delta L_\mathrm{dip}$, where frequency shifts and vector length changes are with reference to calculations at $P_\mathrm{d}=-34$~dBm. Our model shows that $\Delta f_\mathrm{dip}$ is proportional to $\Delta L_\mathrm{dip}^2$. This behavior is consistent with the quartic relation between the frequency of a vibrational mode and the vibrational amplitude caused by weak conservative nonlinearities in the case where the mode is driven by a single tone drive \cite{Carr1999,Husain2003,Kozinsky2006}. It is also consistent with the lineshape broadening of the mechanical response induced by the interplay between conservative nonlinearities and a force noise \cite{Dykman1971}. There, vibrational amplitude fluctuations are converted into resonant frequency fluctuations. If the characteristic amplitude of frequency fluctuations is larger than the linear decay rate of the resonator, the measured response is a superposition of instantaneous responses at various resonant frequencies. As a result, the frequency linewidth of the measured response is proportional to the variance of the amplitude fluctuations \cite{Maillet2017,Huang2019}.
\newpage
\begin{figure}[H]
\centering
\includegraphics{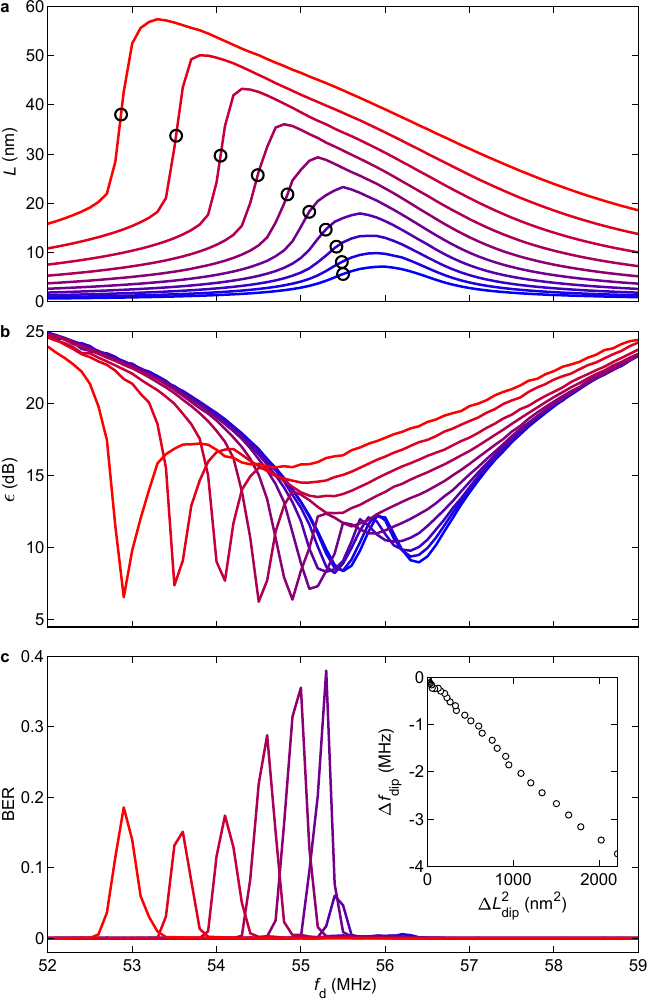}
\caption{Dip in $\varepsilon$ and peak in BER in the nonlinear regime. (a)  $L(f_\mathrm{d})$. Circles indicate the value of $L$ at $f_\mathrm{d}$ where $\varepsilon$ dips. (b) $\varepsilon(f_\mathrm{d})$. (c) $\mathrm{BER}(f_\mathrm{d})$. In (a-c), $P_\mathrm{d}$ ranges from $-31$~dBm (lowest trace in blue) to $-4$~dBm (leftmost trace in red).  Inset to (c): frequency shift $\Delta f_\mathrm{dip}$ of the dip in $\varepsilon(f_\mathrm{d})$ as a function of the square of the change in vector length $\Delta L_\mathrm{dip}^2$, where frequency shifts and vector length changes are with reference to calculations at $P_\mathrm{d}=-34$~dBm. $P_\mathrm{d}$ ranges from $-34$~dBm ($\Delta L_\mathrm{dip}=0$) to $0$~dBm. The effective mass and the coefficients of the conservative and dissipative forces are taken from Supplementary Note~\ref{coefficients}. No addititive white Gaussian noise is used. $R_\mathrm{sym}=10^6$~Hz in all panels.}\label{nonlinearSI}
\end{figure}
\newpage

\subsection{Calculated $L$, $\varepsilon$, $BER$ and constellation diagram vs $f_\mathrm{d}$}

The animated Supplementary Fig.~\ref{anim} shows the calculated $I-Q$ vector length $L$ (top left) and the reciprocal of the error vector magnitude (bottom left) of $I_\mathrm{out}$ and $Q_\mathrm{out}$ as a function of drive frequency $f_\mathrm{d}$. Also shown is the bit error ratio BER as a function of $f_\mathrm{d}$ (bottom right). We use a drive waveform with $25,000$ symbols, a drive power $P_\mathrm{d}=-25$~dBm (Supplementary Note~\ref{section_waveforms}), the coefficients of the conservative and dissipative forces estimated in Supplementary Note~\ref{coefficients}, and the symbol rate $R_\mathrm{sym}=0.5\times10^6$~Hz. Only 5000 symbols are displayed in the constellation diagram (top right). The baseband components of the driving waveform, $\tilde{I}_\mathrm{in}(t)$ and $\tilde{Q}_\mathrm{in}(t)$, are filtered with a root raised cosine (RRC) filter ($\beta=0.35$). Another, identical RRC filter is used at the receiver to obtain $\tilde{I}_\mathrm{out}(t)$ and $\tilde{Q}_\mathrm{out}(t)$ before decimation. Here, we use a receiver output noise of variance $3.9\times10^{-19}$m$^2$ as in the main text.
\newpage
\begin{figure}[H]
\centering
\animategraphics[controls, loop, width=140mm,keepaspectratio]{10}{pR0p5MHz_Pm25dBm}{1}{141}
\caption{Calculated $L$, $\varepsilon$, $BER$ and constellation diagram upon changing $f_\mathrm{d}$. $P_\mathrm{d}=-25$~dBm, $R_\mathrm{sym}=0.5\times10^6$~Hz. 5000 symbols in the constellation diagram are shown out of the $25,000$ symbols used in the calculation. The animation can be controlled with the buttons below the figure.}\label{anim}
\end{figure}

\subsection{Bit error ratio and reciprocal of error vector magnitude for different symbol rates and drive powers}

\begin{figure}[H]
\centering
\includegraphics{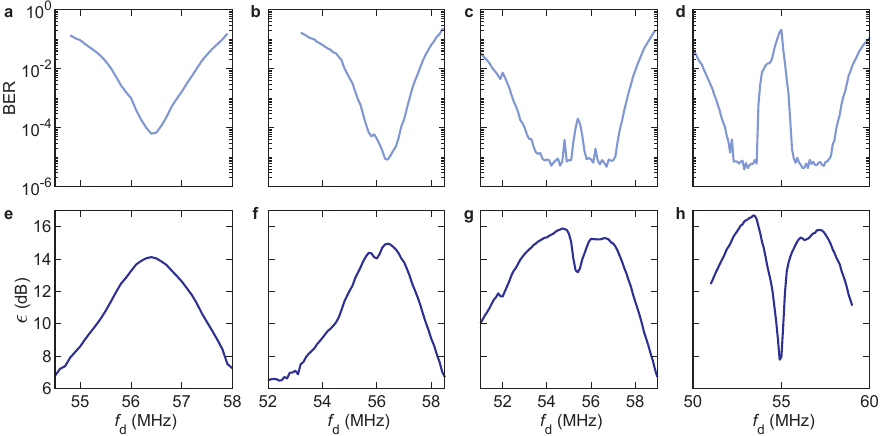}
\caption{BER and $\varepsilon$ as a function of $f_\mathrm{d}$ for $R_\mathrm{sym}=0.5\times10^6$~Hz. $P_\mathrm{d}=-28$~dBm (a, e); $P_\mathrm{d}=-25$~dBm (b, f); $P_\mathrm{d}=-22$~dBm (c, g); and $P_\mathrm{d}=-19$~dBm (d, h).}\label{Chengdu0p5MHz}
\end{figure}

\begin{figure}[H]
\centering
\includegraphics{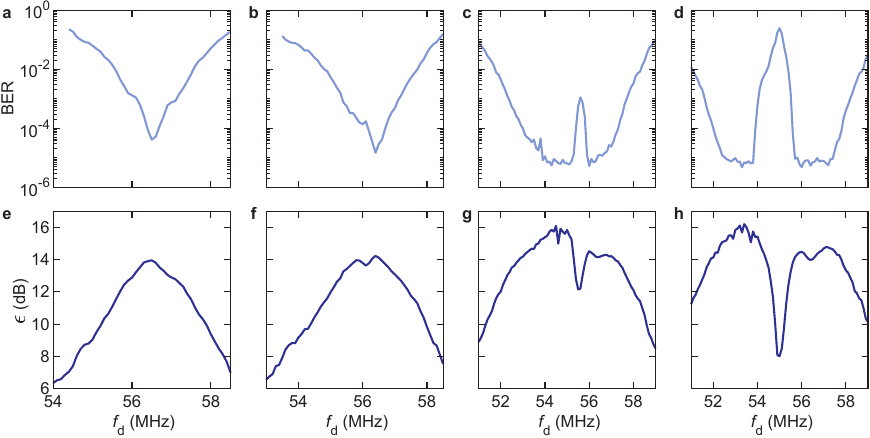}
\caption{BER and $\varepsilon$ as a function of $f_\mathrm{d}$ for $R_\mathrm{sym}=0.75\times10^6$~Hz. $P_\mathrm{d}=-28$~dBm (a, e); $P_\mathrm{d}=-25$~dBm (b, f); $P_\mathrm{d}=-22$~dBm (c, g); and $P_\mathrm{d}=-19$~dBm (d, h).}\label{Chengdu0p75MHz}
\end{figure}

\begin{figure}[H]
\centering
\includegraphics{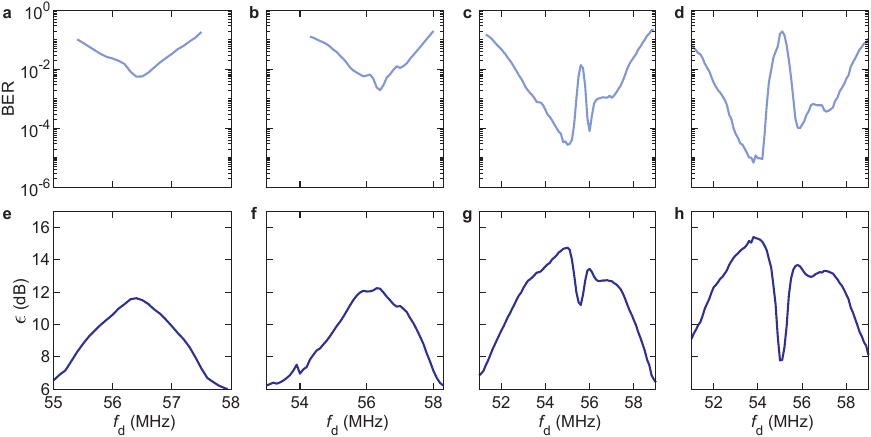}
\caption{BER and $\varepsilon$ as a function of $f_\mathrm{d}$ for $R_\mathrm{sym}=10^6$~Hz. $P_\mathrm{d}=-28$~dBm (a, e); $P_\mathrm{d}=-25$~dBm (b, f); $P_\mathrm{d}=-22$~dBm (c, g); and $P_\mathrm{d}=-19$~dBm (d, h).}\label{Chengdu1p0MHz}
\end{figure}

\begin{figure}[H]
\centering
\includegraphics{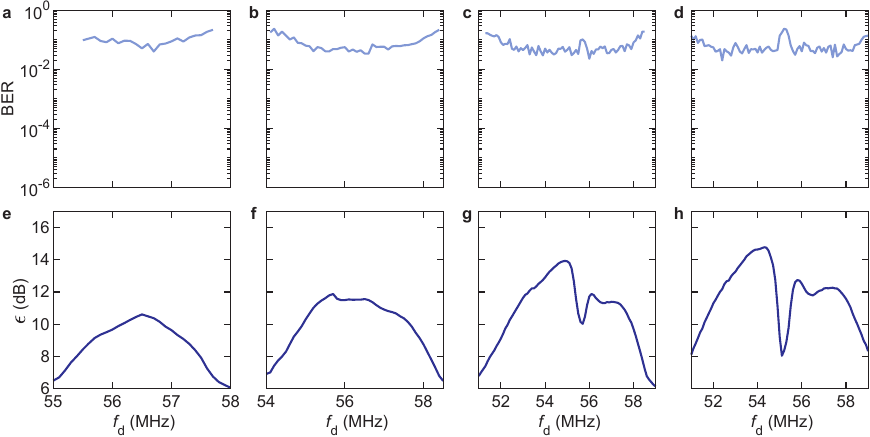}
\caption{BER and $\varepsilon$ as a function of $f_\mathrm{d}$ for $R_\mathrm{sym}=1.25\times10^6$~Hz. $P_\mathrm{d}=-28$~dBm (a, e); $P_\mathrm{d}=-25$~dBm (b, f); $P_\mathrm{d}=-22$~dBm (c, g); and $P_\mathrm{d}=-19$~dBm (d, h).}\label{Chengdu1p25MHz}
\end{figure}

\begin{figure}[H]
\centering
\includegraphics{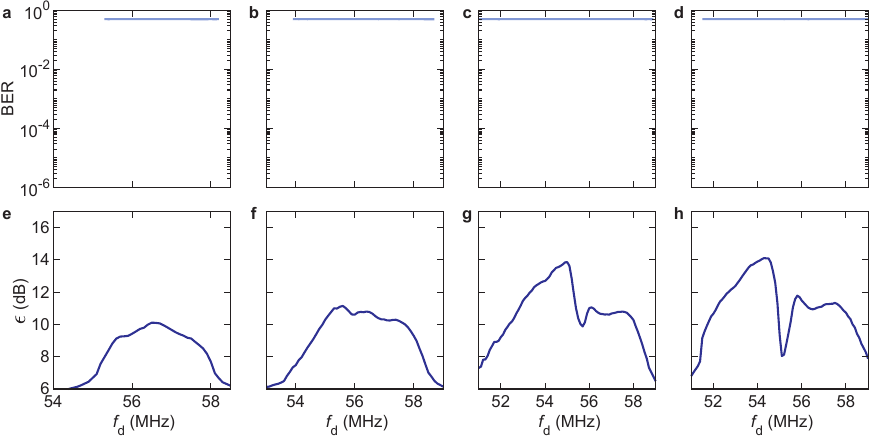}
\caption{BER and $\varepsilon$ as a function of $f_\mathrm{d}$ for $R_\mathrm{sym}=1.5\times10^6$~Hz. $P_\mathrm{d}=-28$~dBm (a, e); $P_\mathrm{d}=-25$~dBm (b, f); $P_\mathrm{d}=-22$~dBm (c, g); and $P_\mathrm{d}=-19$~dBm (d, h).}\label{Chengdu1p5MHz}
\end{figure}

Supplementary Figs.~\ref{Chengdu0p5MHz}-\ref{Chengdu1p5MHz} show the bit error ratio (BER) and the reciprocal of the error vector magnitude as a function of drive frequency $f_\mathrm{d}$ for different symbol rates $R_\mathrm{sym}$ and drive powers $P_\mathrm{d}$. Symbol rates are as follows. $R_\mathrm{sym}=0.5\times10^6$~Hz in Supplementary Fig.~\ref{Chengdu0p5MHz}; $R_\mathrm{sym}=0.75\times10^6$~Hz in Supplementary Fig.~\ref{Chengdu0p75MHz}; $R_\mathrm{sym}=10^6$~Hz in Supplementary Fig.~\ref{Chengdu1p0MHz}; $R_\mathrm{sym}=1.25\times10^6$~Hz in Supplementary Fig.~\ref{Chengdu1p25MHz}; and $R_\mathrm{sym}=1.5\times10^6$~Hz in Supplementary Fig.~\ref{Chengdu1p5MHz}. In all figures, $P_\mathrm{d}=-28$~dBm (a, e); $P_\mathrm{d}=-25$~dBm (b, f); $P_\mathrm{d}=-22$~dBm (c, g); and $P_\mathrm{d}=-19$~dBm (d, h). In Fig.~4i of the main text, data depicting $\mathrm{BER}_\mathrm{min}(R_\mathrm{sym})$ are extracted from the minimum values of $\mathrm{BER}(f_\mathrm{d})$ traces shown here. These data were obtained in a different laboratory and in a noisier environment than the rest of the data presented in the main text, hence the behaviors of $\mathrm{BER}(f_\mathrm{d})$ and $\varepsilon(f_\mathrm{d})$ shown in this section may somewhat differ from the more extensively studied behaviors reported in the main text.

\end{document}